\documentclass[a4paper,10pt]{article}

 \usepackage{siunitx}
 \usepackage{url}
 \usepackage{multirow}
 \usepackage{bm}
 \usepackage{hyperref}
 \usepackage{color}

\usepackage[utf8]{inputenc}
\usepackage{amsthm}
\usepackage{amsfonts}
\usepackage{amssymb}	
\usepackage{amsmath}
\allowdisplaybreaks
\usepackage{mathtools}
\usepackage[english]{babel}
\usepackage{color}
\usepackage{cite}
\usepackage[numbers]{natbib}
\usepackage{slashed}
\usepackage{enumerate}
\usepackage{graphicx}
\usepackage{systeme}
\usepackage{dsfont}
\usepackage{relsize}
\usepackage[margin=0.90in]{geometry}
\usepackage{float}
\usepackage{subcaption}
\usepackage{tikz}
\usepackage[labelformat=simple]{subcaption}

\usepackage{graphicx}
\usepackage{subcaption}
\usepackage{bmpsize}
\usepackage{epstopdf}
\usepackage{bbm}
\usepackage{xcolor,etoolbox}
\usepackage{titling}
\usepackage{authblk}
\newcommand*\samethanks[1][\value{footnote}]{\footnotemark[#1]}
\usepackage{blindtext,graphicx}
\usepackage[absolute]{textpos}
\usepackage[hang,flushmargin]{footmisc}

\usepackage{xfrac}
\usepackage{nicefrac}
\usepackage{esvect}
\usepackage{cases}
\usepackage{empheq}
\usepackage{stmaryrd}
\newcommand{\mi}{\mathrm{i}} %% roman "i"
\newcommand{\comsol}{\textit{Comsol Multiphysics}\textsuperscript{\textregistered}} %% comsol registered trademark
\newcommand{\matlab}{\textit{Matlab}\textsuperscript{\textregistered}} %% comsol registered trademark
\newcommand{\DivT}{\text{Div}} 
\newcommand{\DivV}{\text{Div}}

\def\dataccess#1{{\vskip5.5pt\noindent \textcolor{black}{\fontsize{9}{11}\selectfont Data Accessibility.}\fontsize{8}{11}\selectfont\enskip #1}}

\def\ack#1{{\vskip5.5pt\noindent \textcolor{black}{\fontsize{9}{11}\selectfont Acknowledgements.}\fontsize{8}{11}\selectfont\enskip #1}}
\def\conflict#1{{\vskip5.5pt\noindent \textcolor{black}{\fontsize{9}{11}\selectfont Conflict of interests.}\fontsize{8}{11}\selectfont\enskip #1}}

\title{\vspace{-2cm}Design and experimental validation of a finite-size labyrinthine metamaterial for vibro-acoustics: enabling upscaling towards large-scale structures}
\author{
Svenja Hermann\thanks{Institute of Structural Mechanics, Statics and Dynamics, TU Dortmund University, August-Schmidt-Str. 8,\\ \indent 44227 Dortmund, Germany} 
\, and \, 
Kévin Billon\thanks{Laboratoire de Tribologie et de Dynamique des Systèmes, École Centrale de Lyon, 69134 Ecully, France} 
\, and \, 
Alina-Michelle Parlak\thanks{Institute for Materials in Civil Engineering, TU Dortmund University, August-Schmidt-Str. 8,\\ \indent 44227 Dortmund, Germany} 
\, and \, 
Jeanette Orlowsky\samethanks[3] 
\, and \, 
Manuel Collet\samethanks[2] 
\, and \, 
Angela Madeo\samethanks[1]}

\thanksmarkseries{arabic}
\date{}

\begin{document}

\maketitle

%%%% Abstract text to be placed here %%%%%%%%%%%%
\vspace*{-20pt}
\begin{abstract}
In this paper, we present the design and experimental validation of a labyrinthine metamaterial for vibro-acoustic applications. 
Based on a 2D unit cell, different designs of finite-size metamaterial specimens in a sandwich configuration including two plates are proposed. The design phase includes an optimisation based on Bloch-Floquet analysis with the aims of maximising the band gap and extruding the specimens in the third dimension while keeping the absorption properties almost unaffected.
By manufacturing and experimentally testing finite-sized specimens, we assess their capacity to mitigate vibrations in vibro-impact tests. The experiments confirm a band gap in the low- to mid-frequency range. 
Numerical models are employed to validate the experiments and to examine additional vibro-acoustic load cases. The metamaterial's performances are compared to benchmark solutions, usually employed for noise and vibration mitigation, showing a comparable efficacy in the band gap region. To eventually improve the metamaterial's performance, we optimise its interaction with the air and test different types of connections between the metamaterial and the homogeneous plates. 
This finally leads to metamaterial samples largely exceeding the benchmark performances in the band gap region and reveal the potential of interfaces for performance optimisation of composed structures. 
\end{abstract}

\textbf{Keywords}: Metamaterial, sandwich structure, vibro-acoustics, shape optimisation, experimental testing, interface conditions

% %%%%%%%%%%%%%% Introduction %%%%%%%%%%%%%%% 
\section{Introduction}
Metamaterials are characterised by a specially designed microstructure endowing them, at a larger scale, with unusual properties that cannot be found in natural materials. 
Materials with a periodically arranged microstructure consisting of repeating unit cells show particular dispersion relations for elastic and acoustic waves. For these materials, also known as phononic crystals~\cite{2021_jimenez}, so-called frequency band gaps appear, corresponding to waves that cannot propagate \cite{2018_gan}.
On the microscopic scale, the band gaps can originate from destructive Bragg interference~\cite{2019_zangeneh} of incident and reflected waves that cancel each other out, but also from multiple local resonators in the structure~\cite{2019_romeroGarcia} for which the energy is trapped in the resonator. Bragg interference leads to broader band gaps while local resonators generate stop bands in narrow frequency ranges~\cite{2020_meng} depending on the mass ratio.\\
The possibility to manipulate elastic and acoustic waves has raised the interest of the scientific and engineering community for acoustic metamaterials and application cases such as acoustic cloaking~\cite{2015_buckmann,2016_ma}, focusing~\cite{2011_zhu} and vibration mitigation~\cite{2017_krushynska,2018_bilal} have been proposed during the last years. In the present paper, the focus is put on the use of band gaps in phononic crystals for vibro-acoustic control in civil engineering, such as residential noise.
The main frequencies of the human voice range from \SI{125}{\hertz} to \SI{1600}{\hertz}~\cite{1998_fassold}, which represents a relatively low frequency range. Consequences of exposure to low-frequency noise are annoyance or headache (see for example~\cite{2019_kumar}), however, shielding living rooms against noise in this frequency range remains a challenging task due to the long wave lengths. In solid walls, the homogeneous materials follow the mass-law (cf~\cite{1998_fassold} for example) according to which the sound transmission increases by increasing thickness or mass density of the material. Both cases are not favourable with regard to  limited building space and the need for lightweight constructions. A common solution is the use of double-wall systems that incorporate porous or fibrous sound absorbers like foams~\cite{1996_bolton} or mineral wool (inorganic) but also cork (organic)~\cite{2020_kumar}. However, the material treatments that are needed to obtain an acceptable performance at low frequencies are very heavy~\cite{2021_jimenez}.\\
Different types of metamaterials can offer a solution to this problem. Resonance absorbers like membrane absorbers~\cite{2010_yang,2011_naify,2016_langfeldt} or spring-mass systems \cite{2021_riess,2019_deMelo,2022_kyaw} have been proposed in the context of vibration mitigation in the low frequency range. This type of absorbers attain good attenuation performances but their effect is limited to a narrow frequency band if not coupled with other resonators or additional absorbing materials. 3D phononic crystals that show large band gaps in a lower frequency range have been presented recently~\cite{2017_dAlessandro,2021_gazzola,2019_elmadih}.\\
In this paper, we introduce a new metamaterial for acoustic applications. In particular:
\vspace{-5pt}
\begin{enumerate}
    \item We present a labyrinthine unit cell giving rise to a metamaterial with a low-frequency band gap that can be used in acoustic applications. The band gap region is broad since the unit cell's design mostly exploits the Bragg-scattering mechanism. This unit cell is an optimised version of the cell presented in~[24], such that it was stable enough to be manufactured and mechanically loaded. The thickness of the beams was set equal to the thickness of the air gap throughout the unit cell and the number of air gaps in the radial direction was reduced. Furthermore, the cell size was increased from \SI{2}{\centi \meter} to \SI{5}{\centi \meter}.
    \item We propose three designs of finite-size specimens based on this unit cell by using different base materials and out-of-plane thicknesses. This design procedure comprises a thickness optimisation based on results obtained with established simulation techniques coupling Bloch-Floquet analysis with finite element (FE) method.
    \item We show specimens that have been manufactured as an outcome of this design procedure as small-scale sandwich structures, composed of the metamaterial sandwiched between two plates.
    \item We introduce an experimental campaign that was carried out to validate the capacity of the designed specimens to mitigate vibrations including a  methodology for obtaining the transmission losses at normal incidence from mechanical testing.
    \item We disclose a strong dependence of the vibro-acoustic properties of the sandwich structure on the interface between the metamaterial and the continuous plates that allows to enhance the performances of the specimens significantly.
    \item We explain how the optimised metamaterial specimens will be used as basic building blocks for the design of large-scale structures, thanks to the use of the reduced relaxed micromorphic model. 
\end{enumerate}

%%%%%%%%%%%%%%  Materials and Methods %%%%%%%%%%%%%%% 
\section{Materials and Methods}

\subsection{Design and manufacturing of the metamaterial structures} 
\subsubsection{Unit cell and context} \label{subsec:unitCell}
\begin{figure}[!h]
             \centering
             \begin{subfigure}[c]{0.31\textwidth}
                 \centering
                  \includegraphics[trim = 0 0 0 0, clip,width=\textwidth]{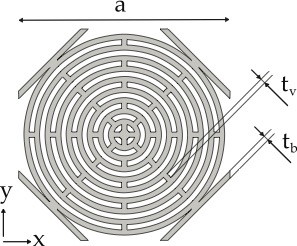}
                 \caption{Unit cell}
                 \label{subfig:experimentalCell}
             \end{subfigure}
             \hfill
             \begin{subfigure}[c]{0.31\textwidth}
                 \centering
                  \includegraphics[trim = 0 0 0 0, clip,width=\textwidth]{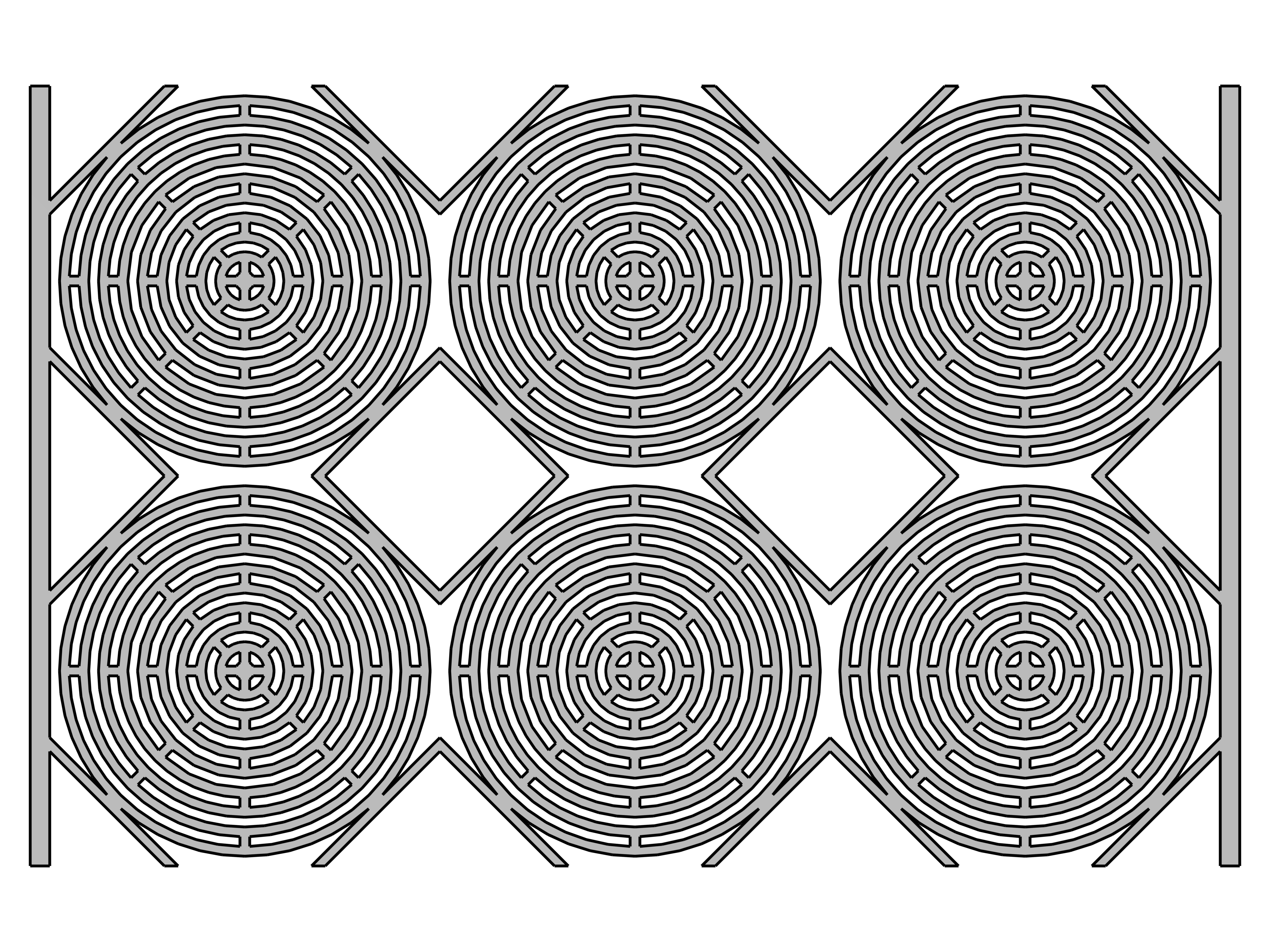}
                 \caption{Array}
                 \label{subfig:array3x3}
             \end{subfigure}
             \hfill
             \begin{subfigure}[c]{0.31\textwidth}
                 \centering
                  \includegraphics[trim = 0 0 0 0, clip,width=\textwidth]{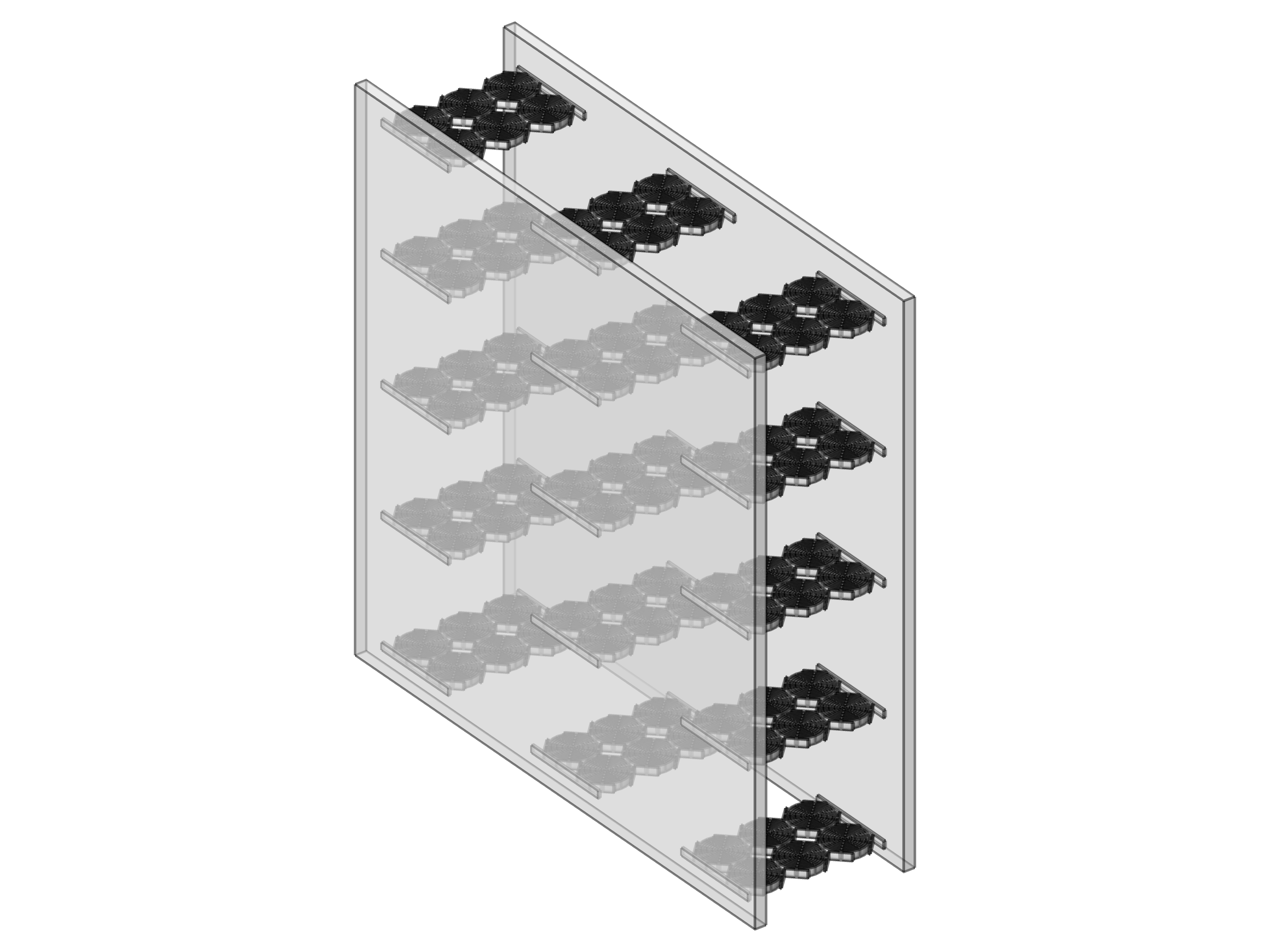}
                 \caption{Panel}
                 \label{subfig:panel}
             \end{subfigure}
             \caption{Labyrinthine metamaterial at different length scales: From unit cell (a) over a cell $2\times3$ array (b) towards a future application scenario, e.g., in an acoustic panel (c). }
             \label{fig:cellInfos}
        \end{figure}

Meta-materials often consist of periodically arranged unit cells. The unit cell studied in this paper has a labyrinthine microstructure (Fig.~\ref{subfig:experimentalCell}), that was developed as a follow-up to the work presented in~\cite{2022_voss}: the cell was further optimised so as to achieve the lowest possible band gap with a characteristic cell size of $a=$~\SI{50}{\milli \meter}. The thickness of the beams in the labyrinth ($t_b$) is \SI{2.5}{\milli \meter} and equal to the thickness of the voids ($t_v$). The ratio of material and void is \SI{43.96}{\percent} in the \SI{50}{\milli \meter}~$\times$~\SI{50}{\milli \meter} square including the unit cell. The prior work showed that the labyrinth unit cell allows to achieve a large band gap. When the cell is manufactured from materials that give rise to low wave speeds (such as polymers) the band gap appears at even lower frequencies in a range that make acoustic wave control an application target. In the present paper, we want to establish the optimal design of a small-scale finite size metamaterial which will be used in future works as a building block for the up-scaling towards larger-scale metamaterial structures (illustration in Figs.~\ref{subfig:array3x3} and~\ref{subfig:panel}). This optimal design of the small scale specimen  will be reached by coupling numerical investigations with experimental validations so as to maximise the absorption properties in the widest possible acoustic frequency range. 

\subsubsection{Design of finite-size metamaterial structures}
\begin{figure}[!bp]
             \centering
             \begin{subfigure}[c]{0.49\textwidth}
                 \centering
                 \includegraphics[trim = 0 0 0 0, clip,width=\textwidth]{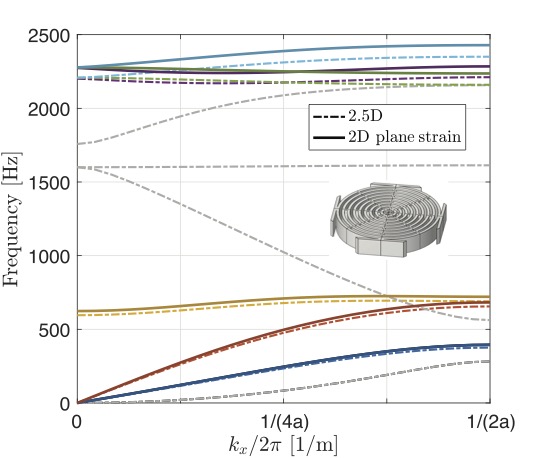}
                 \caption{PMMA unit cell}
                 \label{subfig:dispCurvePlexiCell}         
             \end{subfigure}
             \hfill
             \begin{subfigure}[c]{0.49\textwidth}
                 \centering
                 \includegraphics[trim = 0 0 0 0, clip,width=\textwidth]{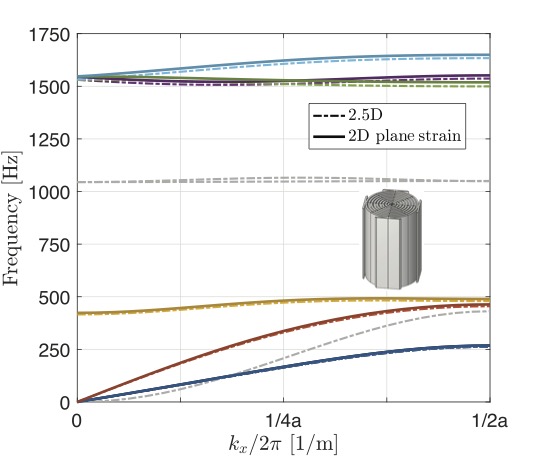}
                 \caption{Photopolymer unit cell}
                 \label{subfig:dispCurvePrintedCell}         
             \end{subfigure}
             \caption{Dispersion diagrams of the labyrinth cells obtained from 2D (plane strain) and 3D Bloch-Floquet simulations for the PMMA unit cell (a) and the optimised photopolymer unit cell (b).}
                \label{fig:dispCurves}
        \end{figure}

In the application case outlined above (see Fig.~\ref{subfig:array3x3}), the metamaterial is clamped between two panels. It is supposed that a vibration of the panels mainly contracts and extends the array. This study therefore concentrates on uni-directional loading of the metamaterial stemming from plane waves in the $x-$direction (cf. Fig.~\ref{subfig:experimentalCell}).
To explore the performance of the labyrinthine cell, the dispersion diagram of an infinitely large array of cells was obtained from a Bloch-Floquet simulation with the FE software \comsol. In the study, the wave propagates in the x-direction and the periodicity conditions are applied on the outer boundaries of the cell. The dispersion diagrams have been obtained for a 2D case with plane-strain approximation and for a case with periodic boundary conditions in the in-plane direction and free boundaries in the out-of-plane direction which is called 2.5D case in the following. No damping was introduced for this first computation. For the convenience of the reader, the implementation of the analysis is detailed in the numerical studies section (Section~\ref{sec:NumSim}). The studies are performed for the two raw materials that will be used later to manufacture the specimens: PMMA and a photopolymer suitable for 3D printing. The raw material's parameters are listed in Tab.~\ref{tab:matProps}. In the 2.5D study, the out-of-plane thickness $t$ of the PMMA cells was determined by the manufacturing constraints and set to \SI{8}{\milli \meter}. The thickness $t$ of the photopolymer cells was set to \SI{73}{\milli \meter}: this value has been determined in an optimisation study with the goal of minimising the frequency range of out-of-plane modes that are invading the band gap (cf. Appendix~\ref{app:BFsimu}). 
The resulting dispersion diagrams for the PMMA and photopolymer arrays are shown in Fig.~\ref{subfig:dispCurvePlexiCell} and Fig.~\ref{subfig:dispCurvePrintedCell} respectively. 
For both unit cells, there are six dispersion curves in the 2D plane-strain case and additional curves for the 2.5D case corresponding to the additional out-of-plane modes. 
In the 2D case, large band gaps can be observed for both, the PMMA (\SI{725}{\hertz}~--~\SI{2237}{\hertz}) and the photopolymer unit cell (\SI{492}{\hertz}~--~\SI{1522}{\hertz}). The dispersion curves of the out of plane modes reduce the band gap of the PMMA material drastically (\SI{1613}{\hertz}~--~\SI{1758}{\hertz}), while they split the band gap up into two parts for the optimised photopolymer cell (\SI{482}{\hertz}~--~\SI{1045}{\hertz}, \SI{1066}{\hertz}~--~\SI{1499}{\hertz}).
From a qualitative point of view, the 2D dispersion curves of the two metamaterials are similar since the geometry of the unit cell is the same. The quantitative differences stem from the raw material properties and, for the 2.5D case, from the out-of-plane thickness $t$ of the cell which influence the propagation of the wave. The comparison of the 2D and 2.5D case for both materials shows that the dispersion curves of the same modes are more similar for the photopolymer cells. This is due to the thickness of the cell which is more similar to a plane-strain case for the photopolymer cell (\SI{73}{\milli \meter}) than for the PMMA cell (\SI{8}{\milli \meter}). Simulations with increasing thickness show that the dispersion curves of the in-plane modes of the 2.5D case converge do the 2D plane-strain solution (Appendix~\ref{app:Converge}). 
In the optimal scenario, the out-of-plane modes are not excited in the in-plane loading that is imagined for the future application. However, due to variations of the material properties and tolerances in the manufacturing, the 3D modes that invade the band gaps have to be considered for the testing of the metamaterials to check that this hypothesis is verified.

\begin{figure}[!h]
             \centering
             \begin{subfigure}[c]{0.49\textwidth}
                 \centering
                 \includegraphics[trim = 0 0 460 0, clip,height = 4cm]{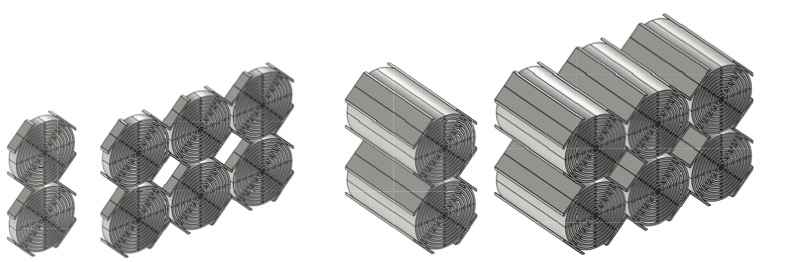}
                 \caption{PMMA metamaterial structures}
                 \label{subfig:fitnieSpecs2Plex}         
             \end{subfigure}
             \hfill
             \begin{subfigure}[c]{0.49\textwidth}
                 \centering                 
                 \includegraphics[trim = 310 0 0 0, clip,height = 3.75cm]{Figures/fitnieSpecs2-eps-converted-to}
                 \label{subfig:fitnieSpecs2Print}  
                 \caption{Photopolymer metamaterial structures}
             \end{subfigure}
             \caption{Finite-size metamaterials comprising $1\times2$ and $3\times2$ unit cells, respectively.}
                \label{fig:specsFinite}
        \end{figure}

In order to manufacture the designed specimens, a finite number of unit cells had to be chosen. On the one hand, the vibration attenuation capacity of the metamaterial increases with the number of unit cells in the direction of wave propagation. On the other hand, this number should remain small enough to obtain a specimen that is easy to handle in the experiments and, in addition, the number should remain small to build an acoustic panel with reasonable thickness and affordable manufacturing costs. In view of the competing interests, numerical simulations of finite specimens with increasing number of unit cells (from 1 to 9) in the direction of wave propagation have been performed. For the in-plane direction perpendicular to the direction of wave propagation a number of 2 cells was chosen due to manufacturing constraints e.g., the size limitations of the 3D printer. Details about this study can be found in Section~\ref{sec:results}. A structure with $3\times2$ unit cells proves to be a good compromise to obtain a measurable effect of the band gap while keeping the dimensions of the specimen reasonable for a first experimental test. Half of the experimental specimens therefore consist of metamaterial arrays with $3\times2$ unit cells. The second half of the specimens possesses $1\times2$ unit cells (cf.~Fig.~\ref{fig:specsFinite}), the band gap effect should be much less present in these specimens. Manufacturing metamaterials with a different number of unit cells will allow us to assess the metamaterials' performances by comparing their responses.

\subsubsection{Manufacturing of finite-size specimens}
\begin{figure}[!h]
    \centering
    \begin{subfigure}[c]{0.49\textwidth}
          \centering   
          \includegraphics[trim = 0 0 0 0, clip,width=\textwidth]{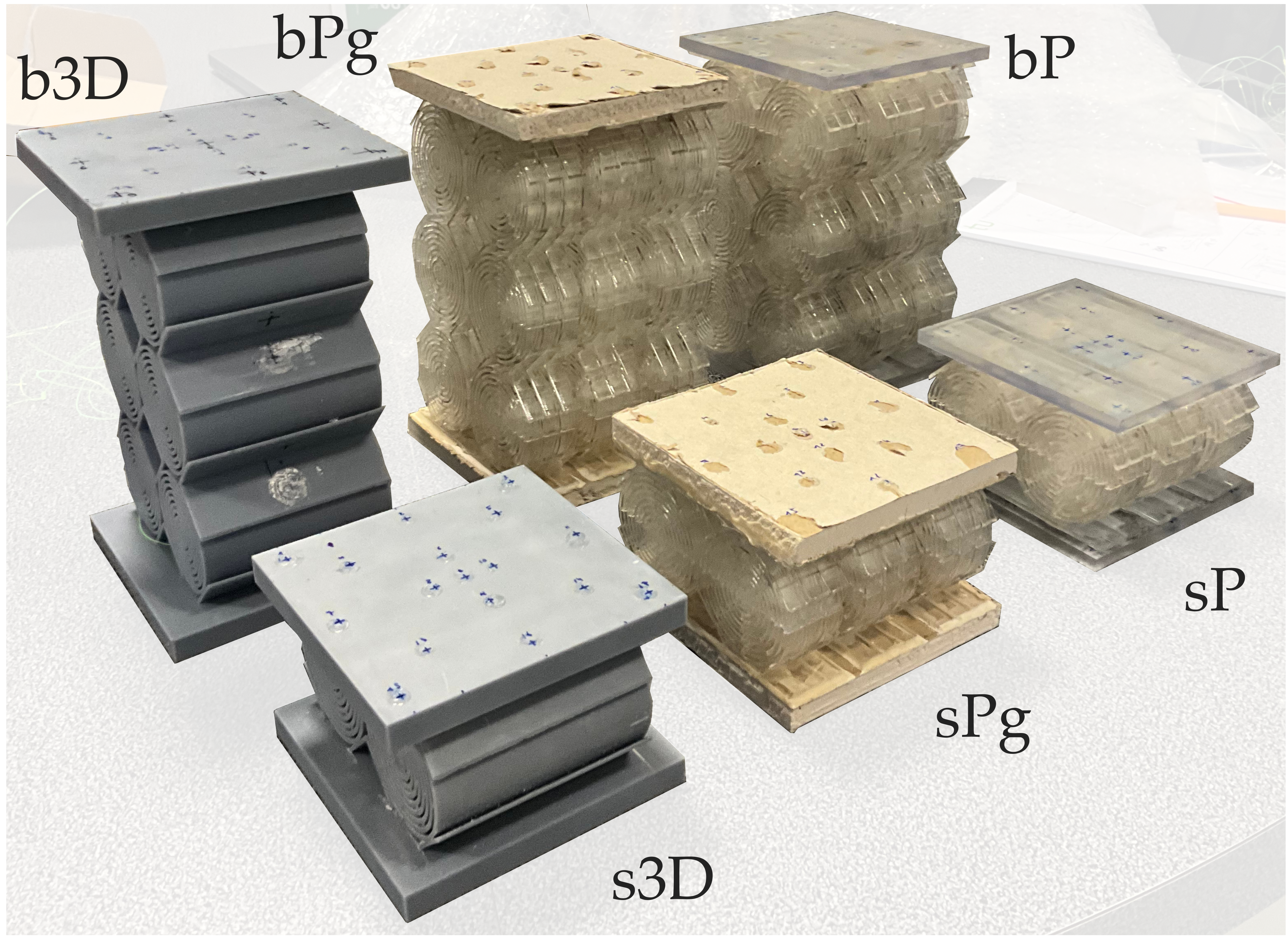}
          \caption{All specimens (after testing)}
          \label{subfig:allSpecsTransparent}         
    \end{subfigure}
    \hfill    
    \begin{subfigure}[c]{0.24\textwidth}
        \centering                 
        \includegraphics[trim = 0 0 0 0, clip,width=\textwidth]{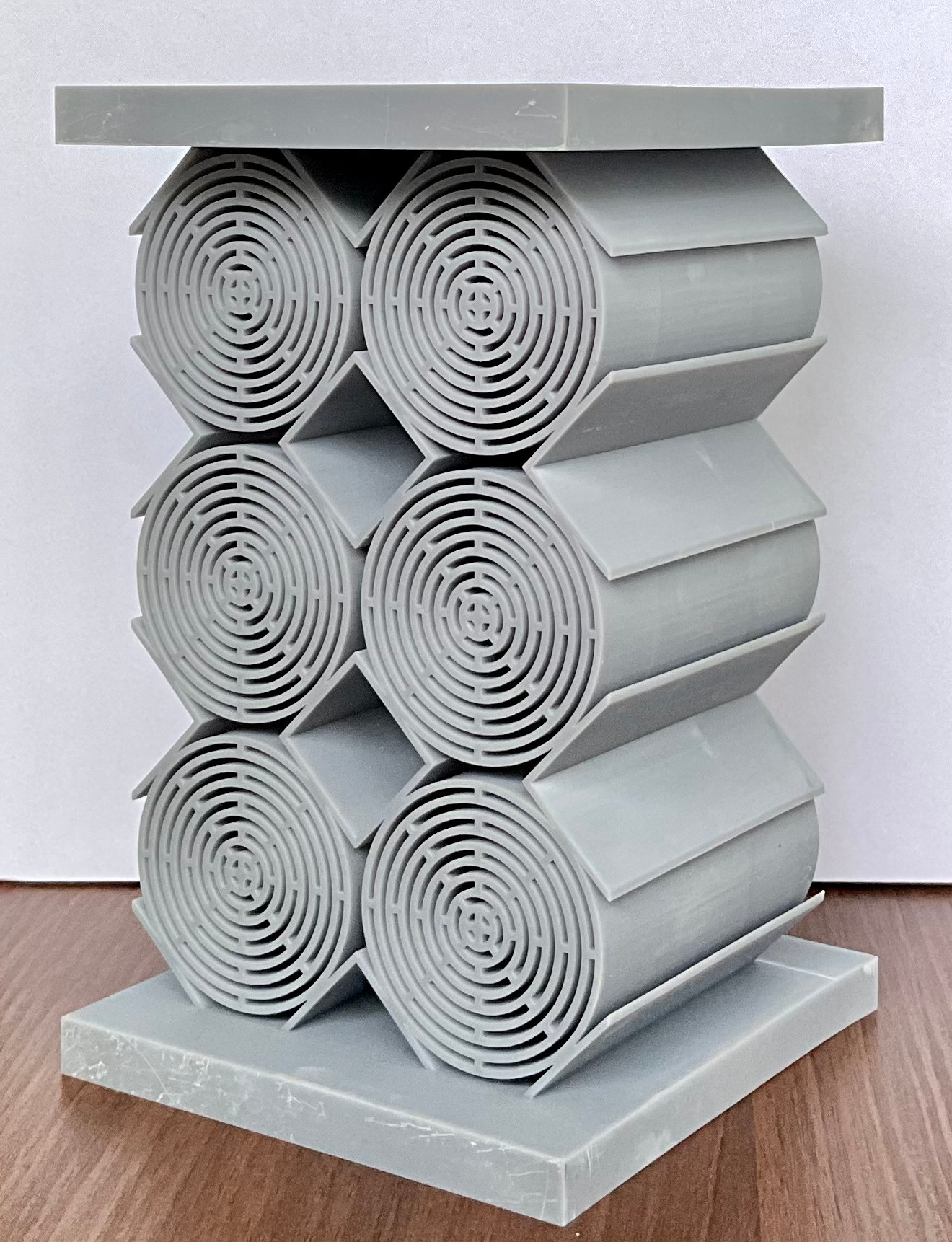}
        \caption{b3D}
        \label{subfig:bigPrinted}  
    \end{subfigure}
    \hfill
    \begin{subfigure}[c]{0.19\textwidth}
        \centering                 
        \includegraphics[trim = 0 0 0 0, clip,width=\textwidth]{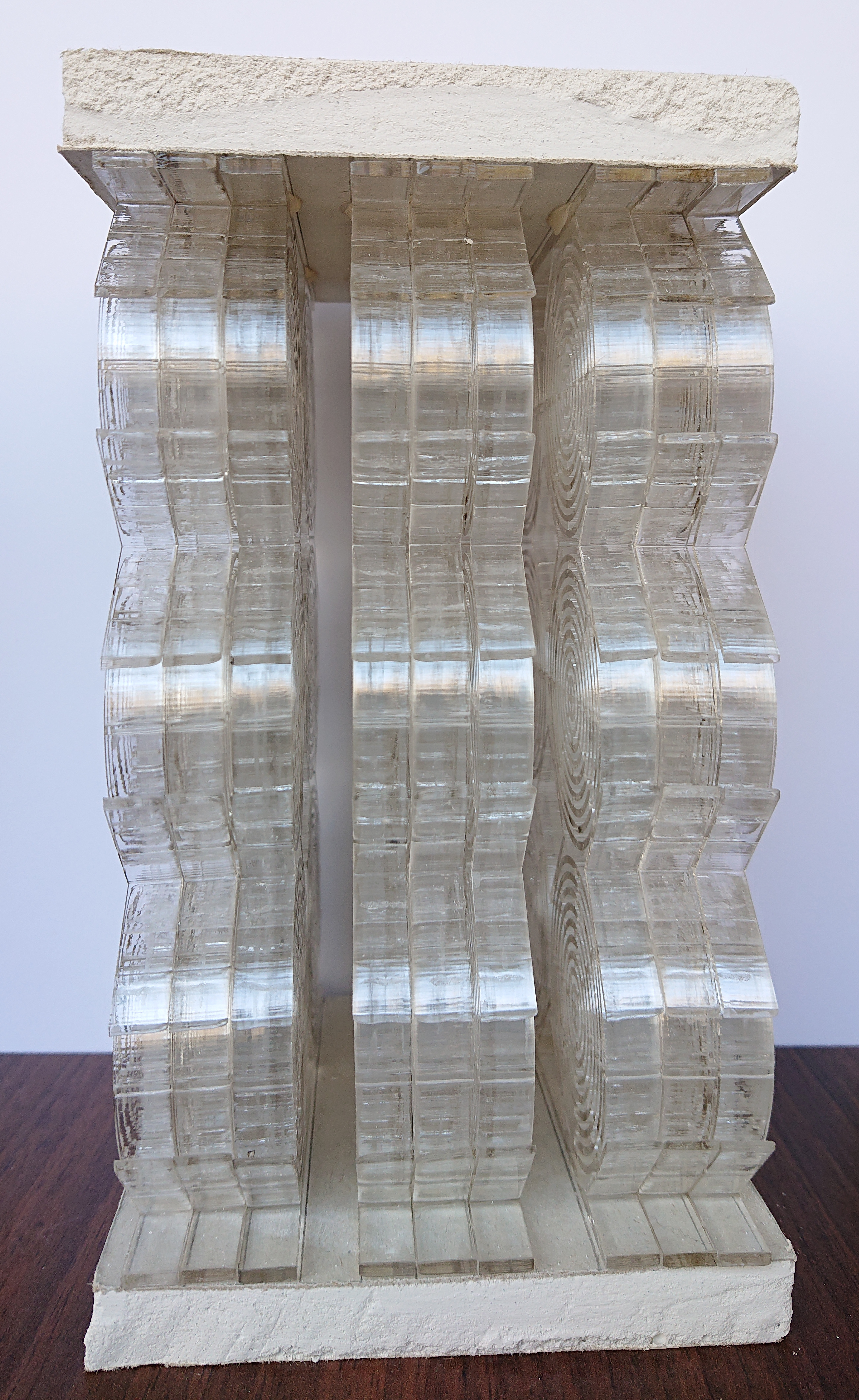}
        \caption{bPg}
        \label{subfig:bigGypsumSide2}  
    \end{subfigure}
    \caption{Metamaterials' specimens: all specimens from left to right: big 3D printed specimen (b3D), small 3D printed specimen (s3D), big PMMA + gypsum specimen (bPG), small PMMA + gypsum specimen (sPG), big PMMA specimen (bP), small PMMA specimen (sP) (a) and single specimens b3D in perspective view (b) and sPg in side view (c).}
    \label{fig:allSpecs}
\end{figure}

Based on the design study outcomes, three different sets of specimens were manufactured in the $3\times2$ and $1\times2$ configuration respectively. A picture of all experimental specimens is shown in Fig.~\ref{subfig:allSpecsTransparent}. The first set consists of PMMA metamaterial and PMMA plates which are glued on both ends of the unit cell arrays. 
The unit cell arrays were obtained by laser cutting from an \SI{8}{\milli \meter} thick PMMA plate. In order to assemble the metamaterial to the plates, an additional foot of \SI{2.5}{\milli \meter} was added to the ends of the arrays. The square PMMA plates have a thickness of \SI{6}{\milli \meter} and a side length of \SI{10.4}{\centi \meter}. The metamaterial arrays are stacked in packages of three in the assembly (cf. Fig.~\ref{subfig:bigGypsumSide2}). The second set is similar to the first one but, instead of PMMA plates, \SI{1.25}{\centi \meter} thick plates of a composite gypsum board were used. The specimens with gypsum plates have been established to obtain a comparison with materials that are currently used for interior walls of houses in civil engineering.
%(as gypsum).
In both cases, the plates are glued to the cell arrays with X60 glue. The third set was 3D printed with a photopolymer called "Standard Resin" from Anycubic~\cite{2023_anycubic} via Digital Light Synthesis (DLS). The printing was performed in the direction perpendicular to the cross section of the cell. The cell array has an out-of-plane thickness of \SI{7.3}{\centi \meter} and is centred between the plates. The square plates initially have a side length of \SI{10.4}{\centi \meter}. The plates of the smaller specimen, however, had to be ground down to a side length of \SI{9.6}{\centi \meter}, since the metamaterial was not perfectly centred after the printing. More details on the specimen's geometry can be found in Appendix~\ref{app:GeomDetails}.

\subsubsection{Experimental determination of the raw materials' elastic properties}
\begin{figure}[!h]
    \centering
     \begin{subfigure}[c]{0.63\textwidth}
             \centering
              \includegraphics[trim = 0 0 0 0, clip,width=\textwidth]{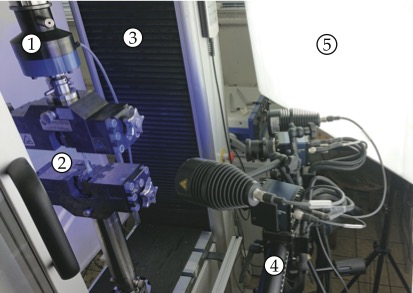}
             \caption{Experimental setup for the tensile tests}
             \label{subfig:tensileTests1}
         \end{subfigure}        
         \hfill
         \begin{subfigure}[c]{0.29\textwidth}
             \centering
              \includegraphics[trim = 0 0 0 0, clip,width=\textwidth]{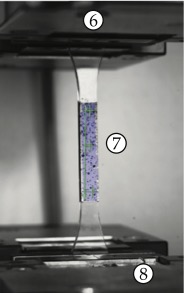}
             \caption{Left camera view}
             \label{subfig:tensileTests2}
         \end{subfigure}
    \caption{Experimental setup of the tensile tests (a) and view on the specimen from the left camera (b). 1~--~Load cell, 2~--~Clamped specimens, 3~--~Universal testing machine, 4~--~Optic measurement system Aramis including 2 cameras and spots, 5~--~Additional lighting, 6~--~Upper clamp, 7~--~Clamped PMMA specimen with evaluation zone, 8~--~Lower clamp}
    \label{fig:tensTests}
\end{figure}

The Young's modulus $E$ and the Poisson ratio $\nu$ of the PMMA and the photopolymer were assessed in uni-axial tensile tests (ISO 527-1, ISO 527-2) in combination with digital image correlation . Therefore, dog bone type specimens were manufactured from PMMA and the photopolymer according to the geometry of specimen type B in ISO 527-2 scaled for a thickness of \SI{3}{\milli \meter}. 
Photos of the experimental setup and of a clamped PMMA specimen are shown in Figs.~\ref{subfig:tensileTests1} and~\ref{subfig:tensileTests2} respectively. 
The mass density $\rho$ of both raw materials was obtained in Pyknometer tests, in which samples of the specimen are weighted in a Pyknometer filled with air and water respectively. The results of the two experimental campaigns are presented in Tab.~\ref{tab:matProps}. The material properties of the gypsum are taken from the data sheet.

\begin{table}[!h]
    \centering
    \caption{Properties of the raw materials used to manufacture the metamaterial specimens.}
    \label{tab:matProps}
    \begin{tabular}{lccc}
        \hline
         Material       & $E$ [GPa] & $\nu$ & $\rho$ [kg/m$^3$]   \\
        \hline
         PMMA           & 3.14 $\pm$ 0.08 & 0.36 $\pm$ 0.02 & 1230 $\pm$ 25 \\
         Photopolymer   & 1.49 $\pm$ 0.12 & 0.42 $\pm$ 0.01 & 1210 $\pm$ 30 \\
         Gypsum         & $\geq$ 2.2     &  0.30            & 680           \\
        \hline
    \end{tabular}
\end{table}
%\vspace*{-4pt}
\newpage

\subsection{Evaluation of the metamaterial's mechanical and acoustic properties}
In the present work, the metamaterial's properties are analysed from both a purely mechanical and a vibro-acoustic (i.e., considering solid air interactions) point of view. The transfer function and the sound transmission loss at normal incidence are chosen as evaluation metrics and are explained in detail in the following. In addition, we present a method that allows to obtain the sound transmission loss from purely mechanical quantities, which reduces the computational effort significantly.

\subsubsection{Mechanical transfer function}
        When the specimen is subjected to a dynamic mechanical loading the vibration propagates in the material. In the frequency domain, the displacement $U$ at location $x_i$ of a structure as response to a point-wise excitation force $F$ (cf. Fig.~\ref{subfig:methodExpA}) at location $x_j$ can be expressed as
          \begin{equation}
             U(x_i,\omega,F) =  \sum_{k=1}^{M} \frac{\Phi_k(x_i) \Phi_k(x_j)}{\omega_i^2 - \omega^2+ 2 \mi \xi \omega \omega_i} F(x_j,\omega) 
             =  Z(x_{i},x_{j},\omega) \, F(x_j,\omega)  \, ,
          \end{equation}
          where $\omega$ is the angular frequency, $M$ is the number of considered eigenvalues $\omega_i$ with the corresponding eigenvectors $\Phi$, $\mi$ is the imaginary unit and $\xi$ is the damping ratio. In the experiments, we measure the acceleration instead of the displacement. In the frequency domain, the displacement and the acceleration $A$ are related by $A(\omega) = (\mi\omega)^2 \, U(\omega)$. The acceleration can, therefore, be written as 
          \begin{equation}
             A(x_i,\omega,F) =  -\omega^2  Z(x_{i},x_{j},\omega) \, F(x_j,\omega)  =  Y(x_{i},x_{j},\omega) \, F(x_j,\omega)  \, .
          \end{equation}
                  
          We obtain the frequency response functions $Y$ for the $N$ tested points from the experiments described in section 2(b) of the manuscript. We want to represent a homogeneous loading originating from a plane pressure wave. Therefore, we assume that the total force $F_t$, which is obtained by integrating the uniform pressure $p$ over the surface $S$ of the input plate, is equally distributed on our $N$ test points such that
          \begin{equation}
              F(\omega) = \frac{p(\omega) S}{N}  = \frac{F_t(\omega)}{N} \; .
          \end{equation}
           The acceleration at a specific location $x_i$ due to the assumed simultaneous loading on the $N$ tested points ($x_j = x_1, ..., x_N$) can be obtained by summing the results obtained for the point-wise loadings. Since $F$ is similar for all points, the summation equals $N F$ and the product can be rewritten such this term is taken out of the summation:  
          \begin{equation}
          \begin{split}
              A(x_i,\omega, F) &= \sum_{j}^{N} A(x_i,\omega, F) =   \sum_{j}^{N} \left(Y(x_{i},x_{j},\omega) \,F(\omega) \right) \\
              &= N F(\omega) \; \sum_{j}^{N} Y(x_{i},x_{j},\omega)  =   F_t(\omega) \, \sum_{j}^{N}  Y(x_{i},x_{j},\omega) \;   .
          \end{split}              
          \end{equation}          
          In the next step, we want to calculate the spatial average of the acceleration which is defined as
          \begin{equation}
                  A(\omega) =  \frac{1}{N} \sum_i^N A(x_i,\omega) 
                  = \frac{ F_t(\omega)}{N} \sum_{i}^{N} \sum_{j}^{N} Y(x_{i},x_{j},\omega) \; .
          \end{equation}     
          The average frequency response function can hence then be obtained from 
          \begin{equation}
             \frac{A(\omega)}{F_t(\omega)} =  \frac{1}{N} \sum_{i}^{N} \sum_{j}^{N} Y(x_{i},x_{j},\omega) = \overline{FRF}(\omega)\; .
             \label{eq:avTF}
          \end{equation}     
          
         We then divide the average frequency response function of the output by the average frequency response function of the input. Since the loading force is similar for both frequency response functions, the force terms cancel each other out (cf. Equation~\eqref{eq:avTF}) and we obtain the ratio of output and input acceleration. The two quantities are related by the transfer function $\overline{TF}$, which can be obtained experimentally from the frequency response functions~$Y$. 
         \begin{equation}
            \frac{\overline{FRF}^{out}}{\overline{FRF}^{in}}
            = \frac{A^{out}(\omega) }{A^{in}(\omega)} 
            = \frac{ \sum_{i}^{N} \sum_{j}^{N} Y^{out}(x_{i},x_{j},\omega)}{ \sum_{i}^{N} \sum_{j}^{N} Y^{in}(x_{i},x_{j},\omega)}
            =\overline{TF}(\omega)
             \label{eq:avTrFun}
         \end{equation}
         Since the transfer function is obtained from spatial averages, it is called average transfer function in the following.

\subsubsection{Sound transmission loss at normal incidence}
\begin{figure}[!h]
    \centering
     \begin{subfigure}[c]{0.3\textwidth}
             \centering
              \includegraphics[trim = 0 0 12 0, clip,width=\textwidth]{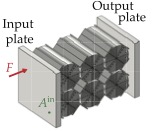}
             \caption{Point-wise loading}
             \label{subfig:methodExpA}
         \end{subfigure}        
         \hfill
         \begin{subfigure}[c]{0.3\textwidth}
             \centering
              \includegraphics[trim = 0 0 12 -10, clip,width=\textwidth]{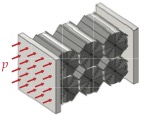}
             \caption{Uniform pressure loading}
             \label{subfig:methodExpB}
         \end{subfigure}  
         \hfill
         \begin{subfigure}[c]{0.35\textwidth}
             \centering
              \includegraphics[trim = 0 -23 0 -23, clip,width=\textwidth]{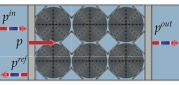}
             \caption{Acoustic loading}
             \label{subfig:methodExpC}
         \end{subfigure}  
    \caption{Illustration of different load cases for a metamaterial specimen: point-like excitation (a), mechanical loading with uniform pressure (b) and acoustic loading with a plane-wave (c). $F$~--~force, $A$~--~acceleration, $p$~--~pressure transmitted to the specimen, $p^{in}$~--~pressure of the incoming plane wave, $p^{ref}$~--~pressure of the reflected plane wave, $p^{out}$~--~pressure of the plane wave radiating from the output plate of the specimen}
    \label{fig:methodExp}
\end{figure}

The sound transmission loss at normal incidence ($TL_{\perp}$) is generally used to indicate the airborne sound insulation of building components. It is a frequency-dependent characteristic, defined by the ratio of the incoming sound power $P^{in}$ and the sound power transmitted by the output plate of the specimen $P^{out}$ as
\begin{equation}
    TL_{\perp}(\omega) = 10 \log_{10} \left( \frac{P^{in}}{P^{out}} \right).
    \label{equation:transmissionLoss}
\end{equation}
In the case of a plane wave, the sound power $P$ can be obtained from the pressure magnitude $|p|$ and the surface $S$ on which the wave is incident according to 
\begin{equation} 
    P = |p|^2 \frac{S}{2 \rho_0 c },
    \label{equation:soundpowerGeneral}
\end{equation}
where $\rho_0$ is the mass density of air and $c$ is the speed of sound in air. 
For an acoustic loading corresponding to a plane wave, is possible to obtain $TL_{\perp}$ from the mechanical quantities stated in the prior section from a uniform loading at the input plate. 
The required expressions for the incident pressure $p^{in}$ and the transmitted pressure on the output plate $p^{out}$ (cf. Fig.~\ref{subfig:methodExpC}) are stated in Equations~\eqref{equation:Pinc} and~\eqref{equation:Ptr}, where $k = \omega / c $ is the wavenumber. The final expression for the sound transmission loss obtained from mechanical quantities is stated in Equation~\eqref{equation:TLgeneral}. The derivation of the different expressions is presented in the Appendix~\ref{app:deriv}. 
\begin{equation}
        p^{in}  =  \frac{\overline{A}^{\, in}}{2} \, \left( \frac{\rho_0}{\mi k} +\frac{1}{\overline{FRF}^{in}  S}  \right),
        \label{equation:Pinc}
\end{equation}
\begin{equation}
        p^{out} =  \overline{A}^{\, out} \frac{\rho_0}{\mi k},
        \label{equation:Ptr}
\end{equation}
\begin{equation}
        TL_{\perp} = 10 \log_{10} \left( \left| \frac{1}{2 \, \overline{TF}} \left( 1+ \frac{\mi k}{\overline{FRF}^{in} S\rho_0}  \right) \right|^2 \right).
    \label{equation:TLgeneral}
\end{equation}

\subsection{Vibro-impact tests on the designed metamaterial structures}
\begin{figure}[!h]
    \centering
         \centering
         \includegraphics[trim = 0 0 0 0, clip,width=\textwidth]{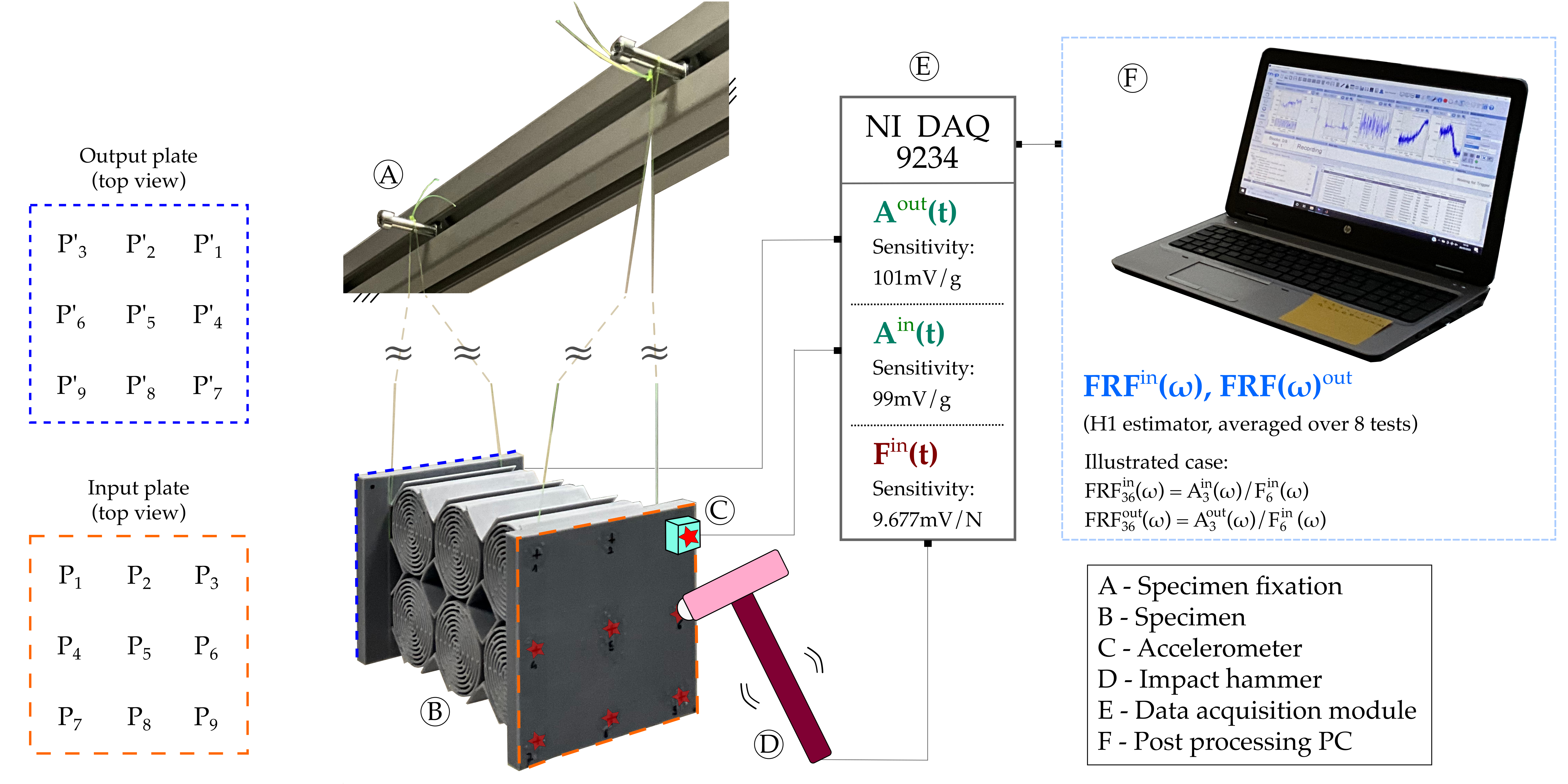}
         \caption{Illustration of the experimental setup. In the represented test case, the structure is excited at point 6 on the top plate and the acceleration is measured at point 3 on the top and bottom plate. The measurement sequence with the accelerometers on point 3 requires eight impacts on each point highlighted with a red star. $a$~--~acceleration, $t$~--~time, $F$~--~mechanical force, $FRF$~--~frequency response function.}
    \label{fig:BCandExpSetup}
\end{figure}

Small specimens are delicate to test since their response is strongly dependent on the boundary conditions. To reduce this influence as much as possible, the specimens were subjected to vibro-impact tests with free boundaries and, on this purpose, suspended from two fishing lines (cf. Fig.~\ref{fig:BCandExpSetup}). Each fishing line was threaded through the hole in the metamaterial that was closest to the connection to the plate. Two accelerometers from PCB were used to measure the acceleration on the input and output plate respectively. The accelerometers were glued to the specimens with an instant adhesive glue from Loctite. For the point-wise excitation, an impact hammer from B\&K (8206-001 58690) with integrated load cell was used. The measurement setup allowed to apply an impact loading on the specimen and to measure the loading force $F$ as well as the acceleration $A$ at one point on the input and one point on the output plate respectively. The signals were recorded and transmitted by a NI 9234 data acquisition module. We used the software M+P for the signal processing. The transfer function was estimated by repeating the same experiment eight times such that the FRF is averaged over the eight tests by a $H_1$ estimator. 
%\textbf{(TODO:verify with Kévin)}. 
The FRF is estimated in a frequency range from \SI{100}{\hertz} to \SI{2}{\kilo \hertz} in steps of \SI{10}{\hertz}. An illustration of the experimental setup is given in Fig.~\ref{fig:BCandExpSetup}. 

Tests of metamaterials with mechanical excitation have successfully been performed in earlier work (cf.~\cite{2016_Dalessandro} for example). In the present work, we test the specimens in a specific configuration allowing us to approximate an acoustic pressure wave with uniform pressure. The goal is to get a first impression of the specimens' vibro-acoustic damping capacity from the experiment. On this purpose, a method to approximate a uniform loading with the experimental setup presented above was developed. 
Several tests are performed to obtain the FRFs on different loading and measurement points $x_j$ and $x_i$. In the present study, nine points are evaluated on each side of the specimen. The points were arranged in a $3\times3$ grid on the plates (cf. Fig.~\ref{fig:BCandExpSetup}) where the outer points had a distance of \SI{1}{\centi \meter} to the closest outer end of the plate and there was a distance of \SI{4.2}{\centi \meter} between the points. The $N$th points on input and output plate, $P_N$ and $P'_N$ respectively, face each other across the specimen. In this configuration, 81 transfer functions are required to obtain the matrix $FRF_{ij}(\omega)$ with $i,j = 1, ... , 9$. Due to reciprocity ($A_i/F_j = A_j/F_i$), only $N(N+1)/2=45$ tests have to be performed. 
At the beginning of a test sequence, the accelerometers are glued on the points $P_1$ and $P'_1$ respectively. The impact loading is applied on points $P_1$ to $P_9$ one after another. After all points have been loaded, the accelerometers are glued to points $P_2$ and $P'_2$. Since $A_2/F_1 = A_1/F_2$, the loading on $P_1$ can be omitted and the loading is applied only to points $P_2 - P_9$. Generally speaking, for a point $P_N$, the loading is applied on points $P_N - P_9$. An illustration for $N=3$ is illustrated in Fig.~\ref{fig:BCandExpSetup}. 
The experimental results are post-processed with \matlab. Therefore, the FRFs of the input and the output plate are arranged in $i\times j \times \omega$ matrices $FRF_{ij}(\omega)$ where $i,j = 1, ..., 9$ corresponds to the point number on the on the specimen. Due to reciprocity of the FRFs, the matrices are symmetric for each frequency:
\begin{equation}
FRF_{ij}(\omega) = 
\begin{pmatrix}
\frac{A_1}{F_1}(\omega) & \frac{A_1}{F_2}(\omega) & \cdots & \frac{A_1}{F_j}(\omega) \\
\frac{A_2}{F_1}(\omega) & \frac{A_2}{F_2}(\omega) & \cdots & \frac{A_2}{F_j}(\omega) \\
\vdots  & \vdots  & \ddots & \vdots  \\
\frac{A_i}{F_1}(\omega) & \frac{A_i}{F_2}(\omega) & \cdots & \frac{A_i}{F_j} (\omega)\\
\end{pmatrix}
= 
\begin{pmatrix}
\frac{A_1}{F_1}(\omega) & \frac{A_1}{F_2}(\omega) & \cdots & \frac{A_1}{F_j}(\omega) \\
\frac{A_1}{F_2}(\omega) & \frac{A_2}{F_2}(\omega) & \cdots & \frac{A_2}{F_j}(\omega) \\
\vdots  & \vdots  & \ddots & \vdots  \\
\frac{A_1}{F_j}(\omega) & \frac{A_2}{F_j}(\omega) & \cdots & \frac{A_i}{F_j} (\omega)\\
\end{pmatrix}
\end{equation}

An approximation of the average frequency response function $\overline{FRF}$ of a plate is obtained by a summation over all accelerations of a single frequency (cf. Eq.~\eqref{eq:avTF}) and the average transfer function $\overline{TF}$ is obtained according to Equation~\eqref{eq:avTrFun}. With this method, a point-wise loading and measurement of the quantities acceleration and force can be used to approximate a case of a uniform loading. The approximation gets better, the higher the number of points, but as the example in Appendix~\ref{app:pointwiseVsAverage} shows, a small number of points is sufficient to achieve a satisfying approximation in the low to mid-frequency range.

%%%%%%%%%%%%%% Numerical Simulations %%%%%%%%%%%%%%% 
\section{Numerical Simulations} \label{sec:NumSim}

\subsection{Bloch-Floquet simulations} \label{subsec:NumSimBF}
The labyrinthine unit cell as depicted in Fig~\ref{subfig:experimentalCell} is implemented in \comsol as 2-dimensional plane-strain case and as 3-dimensional case for the PMMA and photopolymer respectively. The material parameters from the tensile tests are used in a linear elastic material law. The latter as well as the governing equations for the mechanical domain are stated in Equations~\ref{eq:LinElMatLaw} and~\ref{eq:govEqMech} respectively.
The periodicity conditions of the Bloch-Floquet study are applied on the outer surfaces in the $x-z-$plane and in the $y-z-$plane. For two points, located on the surfaces in $y-z-$plane that are located at the points $\bm{r}_0=(x_0,y,z)$ and $\bm{r}_a=(x_0+a,y,z)$, the periodicity is described by 
\begin{equation}
    \bm{u}(\bm{r_a}) = \bm{u}(\bm{r_0}) \, \text{e}^{- \mi \bm{k} (\bm{r}_a-\bm{r}_0)}
\end{equation}
where the wave vector $\bm{k}=(k_x, k_y,k_z)=(N \,\pi/a, 0, 0)$ is applied in the simulation. In the $x-z-$plane, the description is similar for points located at $\bm{r}_0=(x,y_0,z)$ and $\bm{r}_a=(x,y_0+a,z)$. The parameter $N$ is varied from 0 to 1 in 25 steps of 0.04. In the 3D model, the outer boundaries in the $x-y-$plane are not constrained, similar to the experimental case. The case is, therefore, called 2.5D Bloch-Floquet study.

\subsection{Vibro-acoustic model of the specimens} \label{subsec:MechMod}
\begin{figure}[b]
    \centering
     \begin{subfigure}[c]{0.3\textwidth}
         \centering
         \includegraphics[trim = 0 -30 0 -30, clip,width=\textwidth]{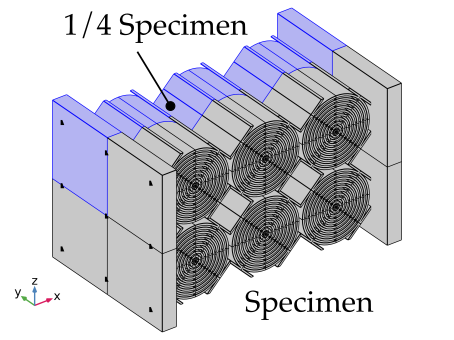}
         \caption{Purely mechanical model}
         \label{subfig:specimenPersp1}         
     \end{subfigure}
     \hfill
     \begin{subfigure}[c]{0.3\textwidth}
         \centering
         \includegraphics[trim = 0 -30 0 -30, clip,width=\textwidth]{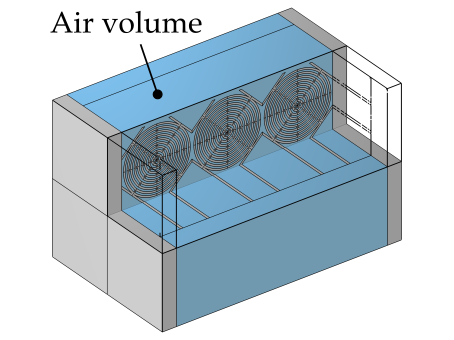}
         \caption{Closed air cavity model}
         \label{subfig:closedAirCav}         
     \end{subfigure}
     \hfill
     \begin{subfigure}[c]{0.36\textwidth}
         \centering
         \includegraphics[trim = 0 0 0 0, clip,width=\textwidth]{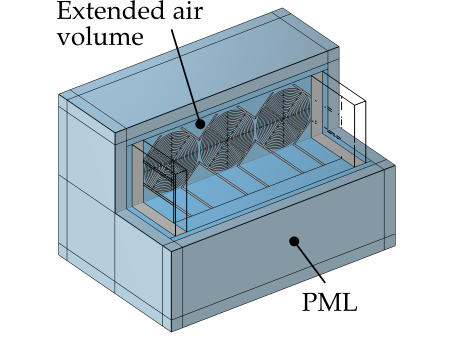}
         \caption{Open air cavity model}
         \label{subfig:openAirCav}         
     \end{subfigure}
    \caption{Implementation of different mechanical/acoustic cases: Purely mechanical model (a), vibro-acoustic model including air in a closed cavity between the plates (b) and an "open air" volume surrounding the entire specimen including the approximation of an infinite domain by a perfectly matched layer around specimen and air volume (c).}
    \label{fig:geomModels}
\end{figure}

A 3D FE model for each of the specimens was implemented in the software \comsol. The model of the printed specimen is shown in Fig.~\ref{subfig:specimenPersp1} as an example. 
For point-like load cases, the simulation was performed for the entire geometry. For cases of uniform pressure loadings, only 1/4 of the specimen can be taken into account when using the appropriate symmetry conditions of zero normal displacement on the surfaces of symmetry, since the specimen and the loading are symmetric. The corresponding quarter of the specimen, implemented in the simulation, is highlighted in  Fig.~\ref{subfig:specimenPersp1}. The implementation shown in Fig.~\ref{subfig:specimenPersp1} was used to assess the purely mechanical performance of the specimens, i.e. the metamaterial's response without air (no mechanical/acoustic coupling). The influence of the surrounding air was taken into account in two vibro-acoustic implementations shown in Figures~\ref{subfig:closedAirCav} and~\ref{subfig:openAirCav} respectively. 
In the first case, a closed air cavity surrounds the metamaterial between the plates and, in the second case, the air is allowed to leave the air cavity and propagate freely. On this purpose, the air volume was increased such as to surround the specimen completely and boundary conditions that approximate an infinitely large air domain were implemented. Therefore, a perfectly matched layer box was situated around the specimen and the air volume.

Purely mechanical studies in the frequency domain with a steady state loading at different frequencies were performed to obtain the frequency response functions similar to the experimental results. The governing equation stems from classical linear elasticity:  
\begin{equation}
 \rho  \omega^2 \bm{u} = \DivT \, \sigma + \bm{f}_v \text{e}^{\mi \phi}  
 \label{eq:govEqMech}
\end{equation}
where \textbf{u} is the displacement vector, Div stands for the divergence operator and the harmonic loading generated by a volume force $\bm{f}_v$ with the phase $\phi$ is represented by $\bm{f}_v$e$^{\mi \phi}$. The isotropic stress tensor $\sigma$ is defined by
\begin{equation}
    \sigma = \left( \frac{E}{1+\nu}  \left( \text{sym} \left(\nabla \bm{u} \right) 
           + \frac{\nu}{(1-2\nu)}  \, \text{tr} \left(\text{sym} \left(\nabla \bm{u} \right) \right) \,  \mathds{1} \right) \right) \left(1+\mi \eta \right)
    \label{eq:LinElMatLaw}
\end{equation}
where $\mathds{1}$ is the identity matrix, $\eta$ corresponds to an isotropic loss factor, introducing complex terms in the stiffness matrix to account for the damping, and $\nabla$ is the gradient operator. 
The material properties ($E$, $\nu$ and $\rho$) and geometric dimensions are chosen such that they correspond to the experimental specimens (cf. Tables~\ref{tab:matProps} and~\ref{tab:specDims} respectively). For the gypsum, a Poisson ratio of $\nu = 0.3$ is assumed. 
The isotropic damping parameter $\eta$ is set by comparison between numerical and experimental measurements and the obtained values are $\eta =$~\SI{4}{\percent} and $\eta =$~\SI{10}{\percent} for the PMMA and the photopolymer respectively. The loss factor commonly introduces a frequency dependence which is, however, not taken into account in the present study.

The air is modelled as an acoustic domain for which a steady state loading is governed by the equation 
\begin{equation}
    \DivV \, \left( -\frac{1}{\rho_0}(\nabla \, p - \bm{q}_d) \right) - \frac{k^2 p}{\rho} = Q_m
\end{equation}
where $k= \omega/c$ is the wavenumber and $\bm{q}_d$ and $Q_m$ are dipole and monopole domain source terms. The properties of the air are its mass density $\rho_0 = $~\SI{1.2043}{\kilo \gram / \meter^3} and the speed of sound $c =$~\SI{343.2}{\meter / \second}.

The strong coupling of the air and the specimen is implemented at the common interfaces with an "Acoustic-Structure-Boundary". On the one hand, the fluid is accelerated by the movement of the structure and, on the other hand, the structure experiences a pressure loading from the air. The boundary conditions are summarised in Equation~\eqref{eq:coupling1} where $\bm{n}$ is the surface normal, $p_t$ is the total acoustic pressure and $\bm{f}_s$ is the surface force per unit area. 
\begin{equation}
    - \bm{n} \cdot \left( -\frac{1}{\rho_0} \left(\nabla \, p - \bm{q}_d \right) \right) = - \bm{n} \cdot \bm{A} \,, \qquad
    \bm{f}_s = p_t \,  \bm{n} \,.
    \label{eq:coupling1}
\end{equation}
Two different load cases are implemented in frequency domain studies in which the frequency ranges from \SI{20}{\hertz} to \SI{2000}{\hertz} in steps of \SI{20}{\hertz}. 
In a first step, a load case similar to the experiment including 9 point-wise loading is implemented to validate the model. The fishing line that suspended the specimens in the experiment is considered to have a negligible influence on the results, the boundaries in the simulation are, thus, left unconstrained. A loading is applied on a small, point-like area and the acceleration are also evaluated point-wise. In order to easily load several points in a parametric study and to reduce the stress concentration, the loading was not implemented as a point load but as a unidirectional surface loading $f_s$ with an intensity specified by the two-dimensional Gaussian function
\begin{equation}
    f_s = \frac{F_0}{S_{plate}} \cdot exp \left(- \left( \frac{(y-y_0)^2}{2 \sigma_y^2}+\frac{(z-z_0)^2}{2 \sigma_z^2} \right) \right),
    \label{eq:pointwLoad}
\end{equation}
where $F_0=$~\SI{1}{\newton} corresponds to the loading force which is distributed over the plate surface $S_{plate}$. The centre of the loading is positioned in the location of the loading points $P=(y_0|z_0)$. The load decreases with the directions $y$ and $z$ with  $\sigma_y=\sigma_z=$~\SI{0.1}{\centi \meter} (distance where the load decreases by \SI{68}{\percent}).
To obtain the average transfer function, 9 simulations with a loading of a single point $P_N$ were performed and the accelerations on the points $P_N - P_9$ were evaluated for each load case. This approach is the opposite of the experimental approach, where the loading was applied on multiple points and the acceleration was measured on a fixed point. The reason is the computational effort: It is faster to vary the loading location in the experiment but for the simulation it is more efficient to reduce the number of point-wise loadings as much as possible. This load case was implemented for the purely mechanical simulation (Fig.~\ref{subfig:specimenPersp1}) and the vibro-acoustic simulation with the open air volume (Fig.~\ref{subfig:openAirCav}).

\begin{figure}[!h]
    \centering
         \centering
         \includegraphics[trim = 0 0 0 0, clip,width=0.9\textwidth]{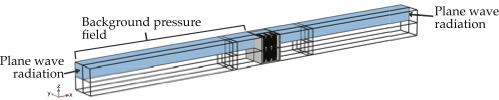}
    \caption{Implementation of an impedance tube model for an acoustic plane-wave loading of the specimen. The scale of the graphic is multiplied by a factor of 0.5 in the direction of the tube axis to enhance the visibility.}
    \label{fig:impTube}
\end{figure}

In the second load case (Fig.~\ref{subfig:closedAirCav}), a uniformly distributed pressure loading is applied to the specimen in two different ways which, however, lead to the same result. Firstly, a mechanical loading with a distributed surface force is used. The force corresponds to Equation~\eqref{eq:pointwLoad}, in which the exponential term is set to 1. This load case was implemented for the three scenarios shown in Fig.~\ref{fig:geomModels}. The calibration of the loss factor $\eta$ is also performed with this load, since this simulation is much less time intensive than the point-wise loading.
%(cf. Tab.~\ref{tab:simCasesInfo}). 
The damping ratio is modified manually, such that the first peak of the transfer function is best fitted. The linear damping only modifies the resonance amplitude but not the frequency.
Secondly, an acoustic loading is implemented in an impedance tube scenario. Therefore, air volumes with the cross sectional area of the plate and a length of \SI{1.5}{\meter} have been added to the simulation (cf. Fig.~\ref{fig:impTube}). A plane wave with normal incidence on the specimen is implemented as "Background Pressure Field" in the first part of the tube and non-reflecting surfaces are defined at the beginning and the end of the tube. 
In the second load case, the specimen is supposed to move as if it was in the impedance tube. Therefore, the displacement of the plates on the outer surfaces was only allowed in the direction of the specimen axis ($x-$~direction). 

% 

%\textbf{Mesh}
The unit cell was meshed first. To obtain a symmetric mesh, the unit cell was divided in eight parts. The first one was meshed with a triangular mesh and the mesh was then copied to the other seven parts. A swept mesh was used in the thickness of the unit cell. For the remaining parts of the model, a swept mesh was used whenever possible, otherwise a tetrahedral mesh was used.
The minimum and maximum possible element size was set to \SI{0.062}{\milli \meter} and \SI{3}{\milli \meter} respectively and Lagrange elements of order 2 were used for the discretisation. The simulations were performed on a desktop computer with a 64-Core processor and \SI{256}{\giga \byte} RAM. The calculation of the transmission loss based on the numerical simulations were validated with a single uniform homogeneous plate, for which the analytical solution is known.
%Information about calculation times $t_{calc}$ and the degrees of freedom (DOF) of the model can be found in Appendix~\ref{app:calcTime}.

\subsection{Reference benchmarks for performance evaluation} \label{subsec:benchmark}
\begin{figure}[!h]
      \centering
      \begin{subfigure}[c]{0.32\textwidth}
        \centering
    \includegraphics[width = \textwidth]{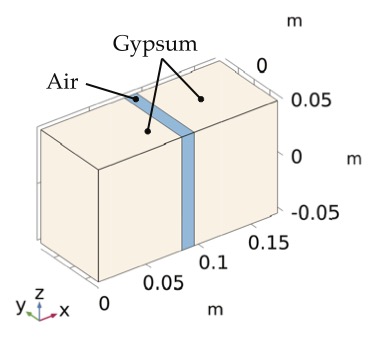}
        \caption{BM1}
        \label{subfig:bmCasesa}
        \end{subfigure}
        \hfill
  \begin{subfigure}[c]{0.32\textwidth}
        \centering
    \includegraphics[width = \textwidth]{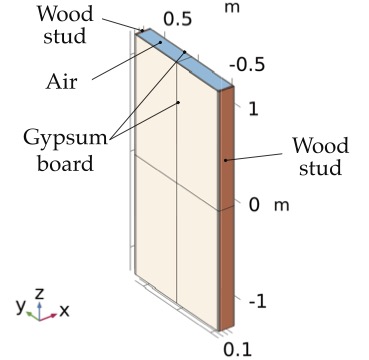}
        \caption{BM2}
        \label{subfig:bmCasesb}
        \end{subfigure}
    \hfill
  \begin{subfigure}[c]{0.32\textwidth}
        \centering
    \includegraphics[width = \textwidth]{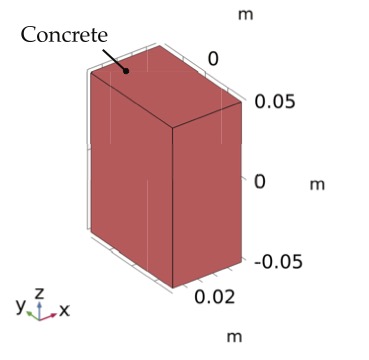}
        \caption{BM3}
        \label{subfig:bmCasesc}
        \end{subfigure}
    \hfill
     \caption{Tested benchmark cases: gypsum plates with air cavity~(a), gypsum boards connected by wood studs~(b) and concrete specimen~(c)}
     \label{fig:bmCases}
\end{figure}

Three reference benchmark cases (BM) are set up to evaluate the performance of the big 3D printed specimen. All three benchmark cases are subjected to a mechanical loading with uniform pressure (load case~2).
%While the first two benchmark cases are used to assess the vibro-acoustic performance of the specimens, the third case is used to evaluate the capacity of the metamaterial to mitigate mechanical vibrations. 
\\
\textbf{BM1}: A double plate configuration with an air cavity between the plates was set up such that the total mass of the two plates included the mass of the metamaterial. Therefore, the mass of the metamaterial was divided between the two plates. We chose gypsum as a material for the plates ($E =$\SI{2800}{\mega \pascal}, $\nu = 0.3$, $\rho = $\SI{680}{\kilo \gram / \meter}$^3$), which is commonly used in civil engineering practice. For each plate, the thickness has been increased from \SI{1}{\centi \meter} to \SI{7.91}{\centi \meter}. The total thickness was kept similar to the thickness of the metamaterial specimen, which leads to a reduction of the air cavity length from \SI{15}{\centi \meter} to \SI{1.18}{\centi \meter} (cf.~Fig.~\ref{subfig:bmCasesa}). 
We included the BM1 to show that the metamaterial structure performs better than this highly idealised case. Indeed, it is clear that a perfect uncoupling between the two plates is not achievable in reality for structural reasons  such as load bearing. To create panels separated by air cavities, the most common solution is to join them together with wooden vertical beams with a certain regular spacing between each other. It is clear that in this case, the panel's performances will be worse than the ideal case, since vibrations will be transmitted through the wooden connecting components.\\
\textbf{BM2}: We performed numerical simulations of a double panel, in which two gypsum plates (thickness = \SI{1}{\centi \meter}) are connected by wood studs. The geometric dimensions of the studs are chosen close to commonly used 2$\times$6 studs for interior walls (width = \SI{3.81}{\centi \meter}), but a thickness of \SI{15}{\centi \meter} instead of \SI{8.89}{\centi \meter} was chosen to maintain the cavity size. The material parameters of construction lumber (pine, spruce, fir) were used ($E =$~\SI{500}{\mega \pascal}, $\nu =$0.23, $\rho =$~\SI{780}{\kilo \gram / \meter^3}). In the chosen configuration, gypsum plates with a common size of \SI{1.25}{\meter}~$\times$~\SI{2.5}{\meter} are tested. The air cavity between the plates measures hence \SI{15}{\centi \meter} $\times$ \SI{121.19}{\centi \meter} $\times$ \SI{250}{\centi \meter}  (cf.~Fig.~\ref{subfig:bmCasesb}).  \\
\textbf{BM3}: We test another case which is frequently employed in civil engineering: a concrete wall. The material parameters for the concrete are the following: $E=$~31.6~GPa, $\rho=$~2275~kg/m$^3$, $\nu=$~0.1, $\eta=$~0.01. For a thickness of \SI{4.61}{\centi \meter}, the mass of the concrete wall is equivalent to the metamaterial mass  (cf.~Fig.~\ref{subfig:bmCasesc}).\\

%%%%%%%%%%%%%% Results %%%%%%%%%%%%%%% 
\section{Results} \label{sec:results}

\subsection{Experimental results and model validation}
\begin{figure}[tp]
    \centering
    \includegraphics[trim = 100 0 115 0, clip,width=\textwidth]{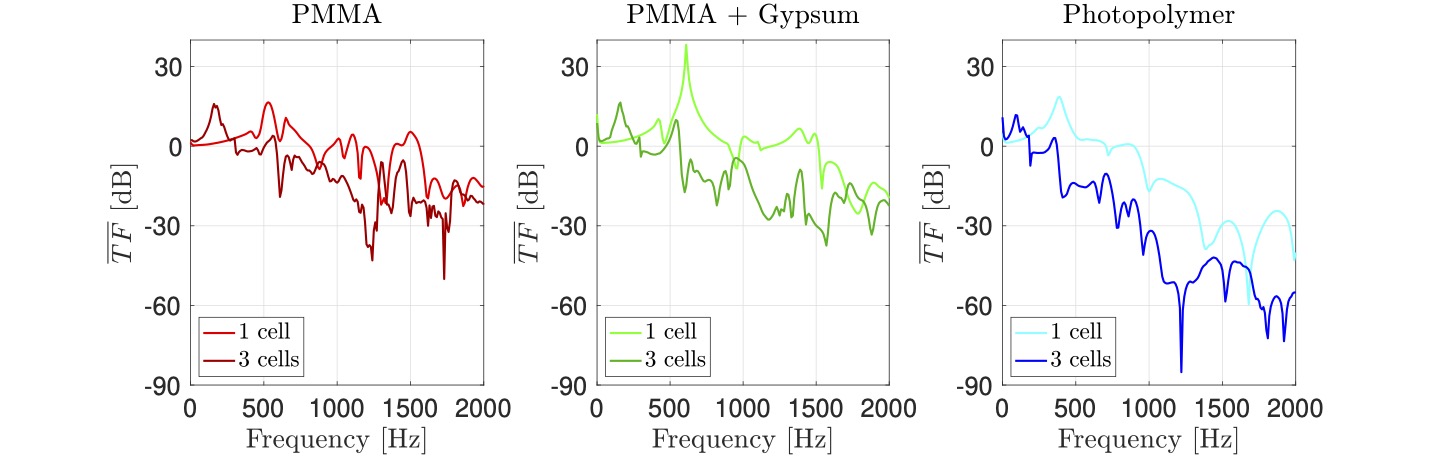}
    \caption{Experimentally obtained average transfer functions obtained for the six specimens.}
    \label{fig:TFmoyen}
\end{figure}

The average transfer functions obtained from the experimental results of the small and big specimens are at first compared for each of the three designs as shown in Fig.~\ref{fig:TFmoyen}. It can be inferred, that the vibration transfer from upper to lower plate is reduced more effectively by the big specimens  (length = 3 cells) than by the small specimens (length = 1 cell) for most of the frequency range. This indicates that a single cell is not enough to obtain the band gap effect. Large drops in the transfer function, which are characteristic for band gaps, can be observed for the big PMMA specimen around \SI{1200}{\hertz} and \SI{1700}{\hertz}. The results for the specimens with PMMA metamaterial and different plates are very similar since they have the same design and the same raw materials are used in the functional part. For the printed specimen, the transfer function drops at \SI{1220}{\hertz}, \SI{1810}{\hertz} and \SI{1920}{\hertz}. The drop at \SI{1220}{\hertz} reaches \SI{-85}{\deci B} which corresponds to a reduction of the vibration amplitude by a factor of approximately 17500.
A comparison of the different designs shows that the optimised design of the photopolymer specimen performs better than the designs that contain the PMMA metamaterials for almost the entire frequency range. In addition to the large drop to approximately \SI{-85}{\deci B}, a major part of the transfer function evolves in a range from \SI{-30}{\deci B} to \SI{-50}{\deci B} which corresponds to a division of the average acceleration by factors of approximately 30 and 300 respectively. Moreover, the first resonance peak of the photopolymer specimen is decreased in intensity and in frequency compared to the PMMA specimens which are both favourable conditions in view of increased acoustic comfort. 
\begin{figure}[bp]
    \centering
    \includegraphics[trim = 98 0 105 0, clip,width=\textwidth]{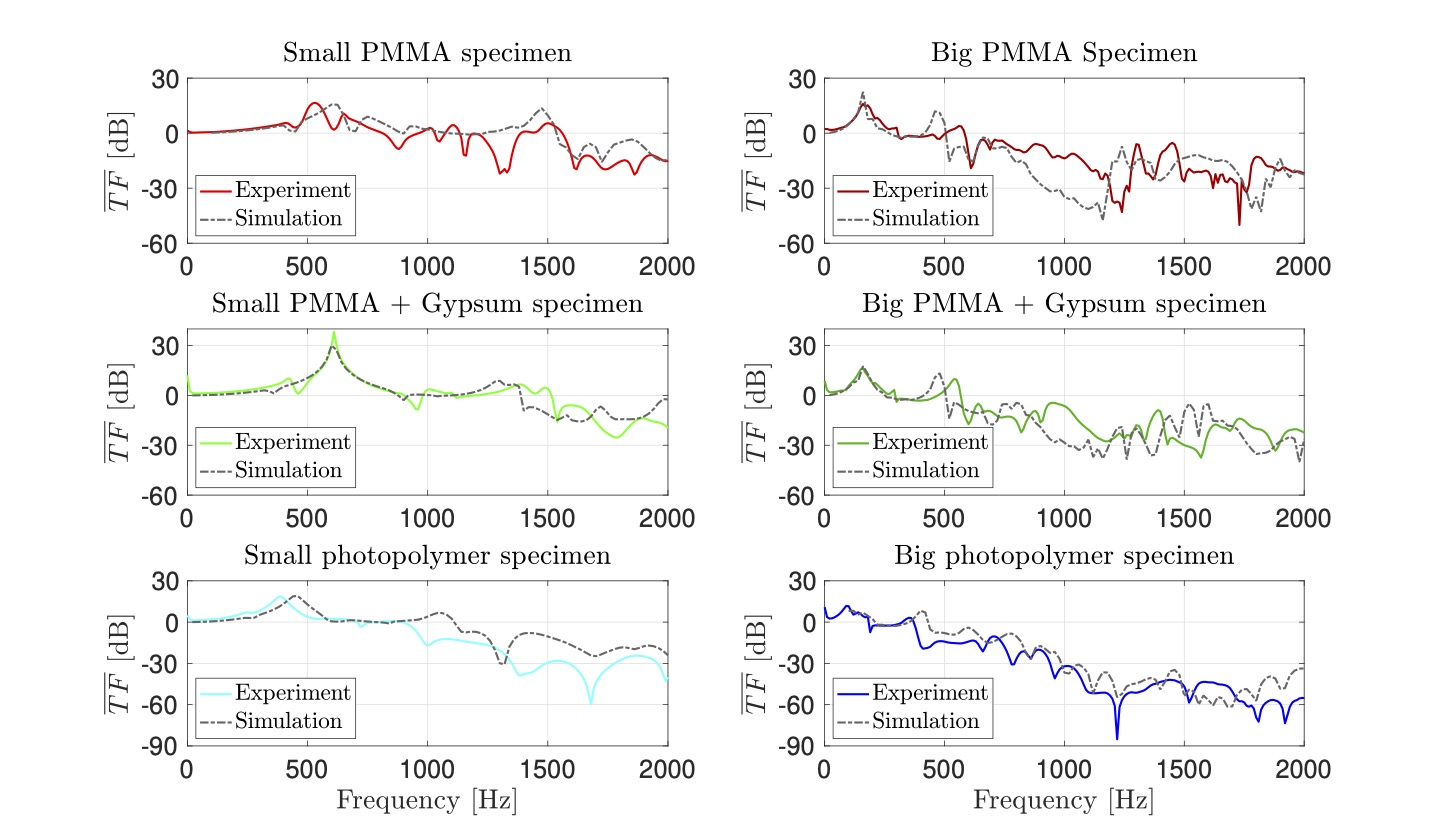}
     \caption{Comparison of the average transfer functions obtained from experiments and numerical simulations for the six tested specimens.}
     \label{fig:exc9ptVsAverage}
\end{figure}
The comparison between the numerical simulation with load case 1 (9-point loading) and experiment for every specimen is shown in Fig.~\ref{fig:exc9ptVsAverage}. Even though the results of the 9-point loading do not show a 100\% agreement with the experimental results, the trend of the curve is captured to a significant extent. Reasons for the disagreement can be defects in the geometry of the specimen parts and in the junction between the metamaterial and the plates. Furthermore, the glue has not been considered in the simulation but has its own material properties and can cause dispersion as well as non-perfect contact. The photopolymer is likely to contain defects that happen during the printing. Nevertheless, from a qualitative point of view the results are in good accordance with the experiment, especially if one considers that only a single calibration process for the linear damping was performed. 
The accordance between simulation and experiment is better for the 3D printed than for the assembled geometries, which confirm that assembly uncertainties may play a considerable role. In addition, the intensity of the vibrations (= captured signals) at high frequency gets weaker at higher frequencies. This, in turn, can be an additional explanation of the discrepancy between experimental and simulation results.

\subsection{Vibro-acoustic properties of the finite-size photopolymer specimen}
\begin{figure}[!h]
      \centering
      \begin{subfigure}[c]{0.28\textwidth}
        \centering
    \includegraphics[trim = 0 0 880 0, clip,height=6cm]{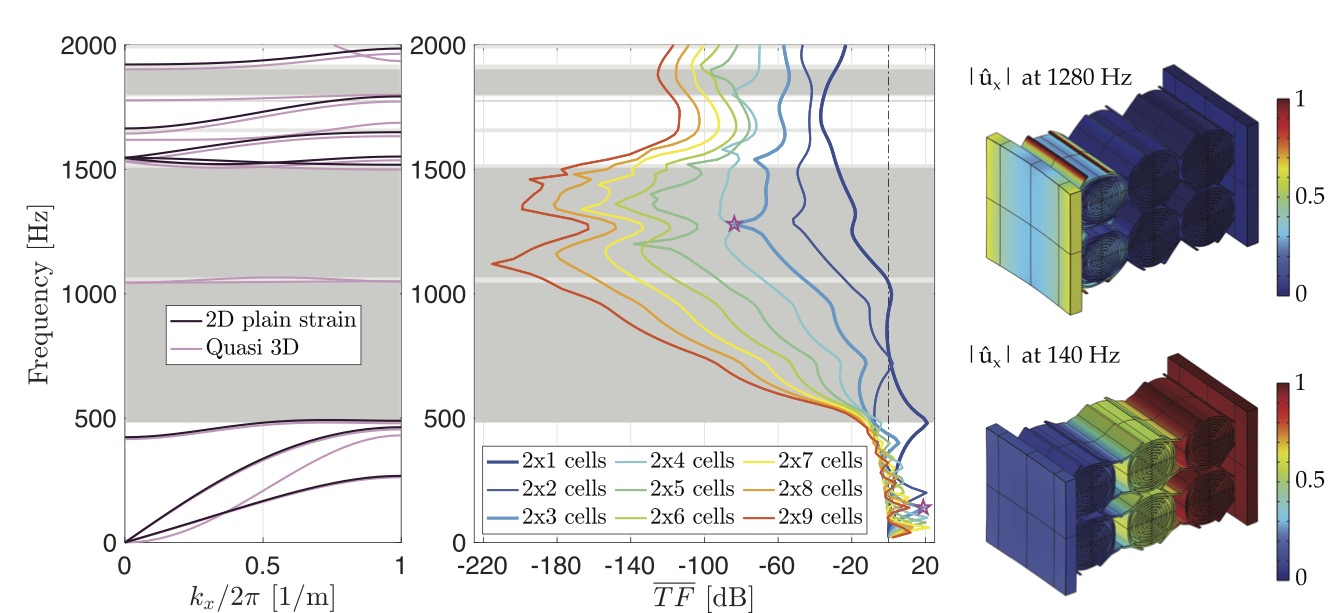}
        \caption{Dispersion curves}
        \label{subfig:simuDispCurvesTFa}
        \end{subfigure}
        \hfill
  \begin{subfigure}[c]{0.42\textwidth}
        \centering
    \includegraphics[trim = 385 0 365 0, clip,height=6cm]{Figures/simuDispCurvesTF-eps-converted-to}
        \caption{Transfer function}
        \label{subfig:simuDispCurvesTFb}
        \end{subfigure}
    \hfill
  \begin{subfigure}[c]{0.28\textwidth}
        \centering
    \includegraphics[trim = 900 0 0 0, clip,height=6cm]{Figures/simuDispCurvesTF-eps-converted-to}
        \caption{Deformed specimen}
        \label{subfig:simuDispCurvesTFc}
        \end{subfigure}
    \hfill
     \caption{Dispersion curves for an infinite photopolymer metamaterial (a), transfer functions for finite photopolymer specimens with different number of unit cells in the $x$~direction (b) and deformed configurations of the experimental design at resonance frequency (\SI{140}{\hertz}) and in the band gap (\SI{1280}{\hertz}) (c). The frequencies of the deformed configurations shown in (c) are highlighted in the transfer function by a pink star in (b). $|u_x|$~--~normalised displacement magnitude in $x$~direction.}
     \label{fig:simuDispCurvesTF}
\end{figure}

The average transfer function of the photopolymer specimen is compared to the dispersion curve in the following to validate the presence of the band gap. Fig.~\ref{subfig:simuDispCurvesTFb} shows the numerical results obtained from load case 2 (average loading) for specimens with increasing number of unit cells. As the number of cells increases, the transfer function drops increasingly in the regions that are supposed to show a band gap according to the dispersion diagram (Fig.~\ref{subfig:simuDispCurvesTFa}). 
Even if the metamaterial with a depth of 9 unit cells shows the best absorption properties, we must acknowledge that the experimental configuration ($2\times3$ cells) provides a vibration reduction between \SI{-20}{\deci B} and \SI{-80}{\deci B} in the band gap region.

The low frequency of the different specimens shows peaks, that correspond to the resonance frequencies of the finite structures. As an example, the deformed configuration of the big photopolymer specimen (2$\times$3 unit cells) at the resonance frequency of \SI{140}{\hertz} is illustrated in Fig.~\ref{subfig:simuDispCurvesTFc}. The colouring corresponds to the normalised displacement in $x$-direction ($|u_x|$) and the corresponding resonance peak is marked with a pink star in the transfer function. The deformed configuration of the specimen in the band gap at \SI{1280}{\hertz} demonstrates that the majority of the vibration remains in the first unit cell of the metamaterial. The vibration mitigation is hence due to the band gap and cannot be achieved with a single unit cell, as the experimental and numerical results show. \\

\begin{figure}[!h]
    \centering
      \begin{subfigure}[c]{0.48\textwidth}
        \centering
    \includegraphics[trim = 0 0 20 0, clip,width=\textwidth]{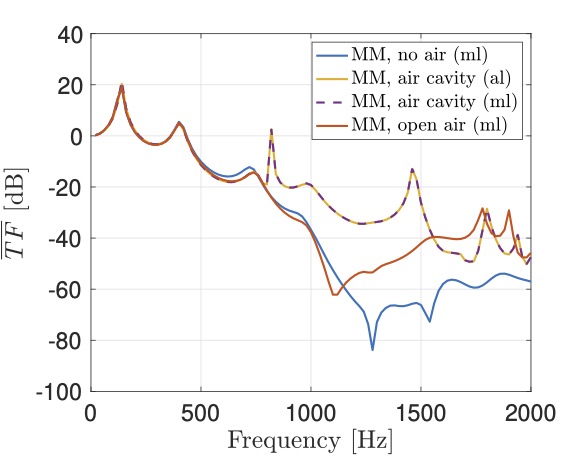}
        \caption{Uniform loading}
        \label{subfig:TFsVibroaca}
        \end{subfigure}
        \hfill
  \begin{subfigure}[c]{0.48\textwidth}
        \centering
    \includegraphics[trim = 0 0 20 0, clip,width=\textwidth]{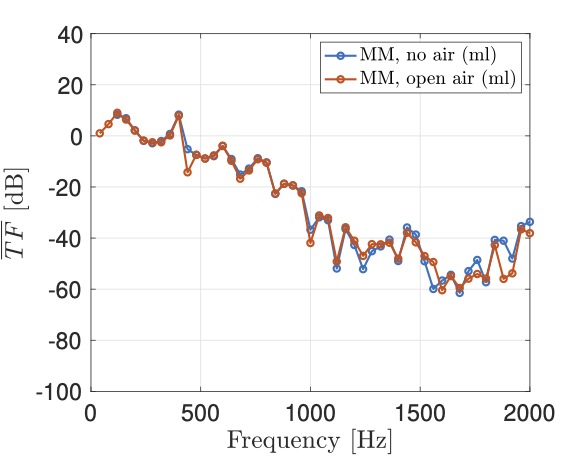}
        \caption{Point-wise loading}
        \label{subfig:TFsVibroacb}
        \end{subfigure}
     \caption{Mechanical transfer functions of the photopolymer specimen obtained from purely mechanical and vibro-acoustic configurations with uniform loading (a) and point-wise loading~(b) respectively. MM~--~metamaterial, ml~--~mechanical loading, al~--~acoustic loading.}
     \label{fig:TFsVibroac}
\end{figure}

After having validated a reasonable effectiveness of the experimental design ($2\times3$ unit cell specimens) with respect to its absorption properties in a purely mechanical situation, we want to understand how the metamaterial's performances can be affected by the presence of air. To do so, the photopolymer specimen was numerically tested in the impedance tube (see Fig.~\ref{fig:impTube}) including the surrounding air in the simulation. Fig.~\ref{subfig:TFsVibroaca} shows the results of the test and also includes the purely mechanical simulation case as well as the mechanical load case with a closed air volume around the specimen. The results show, that the capacity of the specimen to mitigate the vibrations decreases significantly in presence of the closed air volume (case illustrated in Fig.~\ref{subfig:closedAirCav}). An eigenfrequency analysis shows that the sharp peaks in the graphic correspond to acoustic resonances of the system. The graphic also shows, that the acoustic and the mechanical loading lead to the same transfer function, which allows to reduce the simulation time significantly by \SI{30}{\percent}. 
When the specimen is surrounded by a large air volume (c.f. Fig.~\ref{subfig:openAirCav}), the effect of the air is reduced, but the transfer function is still different from the purely mechanical result as Fig.~\ref{subfig:TFsVibroaca} shows. 
This difference between the cases without air and with open air was not expected, since the experimental results are well approximated with a purely mechanical model without air, as it was shown in Fig.~\ref{fig:exc9ptVsAverage}. We thus investigated further the effect of the type of loading and observed that the coupling between the structure and the air does not have a significant effect on the results obtained from the point-wise excitation. The different parts of the plate are not excited in phase when point-wise loads are applied one after another and as a result, the air volume between the plates does not have the same capacity to move as it happens when the plate is excited with a more uniform excitation. This is shown in Fig.~\ref{subfig:TFsVibroacb} in which no significant difference between the case without air and the case of open air can be observed.

\begin{figure}[!h]
    \centering
      \begin{subfigure}[c]{0.48\textwidth}
        \centering
    \includegraphics[trim = 0 0 20 0, clip,width=\textwidth]{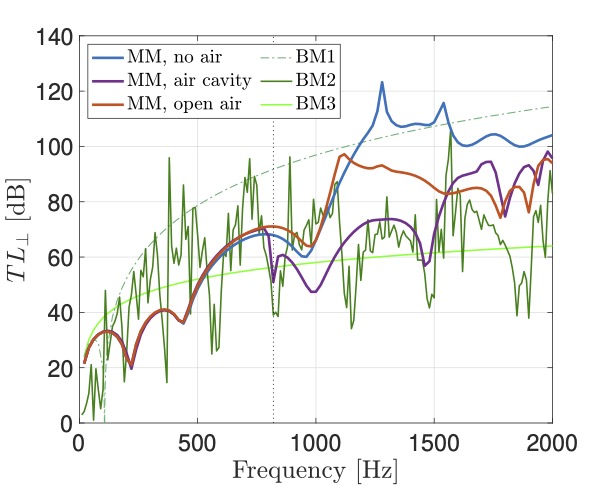}
        \caption{Transmission loss}
        \label{subfig:TLsVibroaca}
        \end{subfigure}
        \hfill
  \begin{subfigure}[c]{0.48\textwidth}
        \centering
        \includegraphics[trim = 0 0 0 0, clip,width=\textwidth]{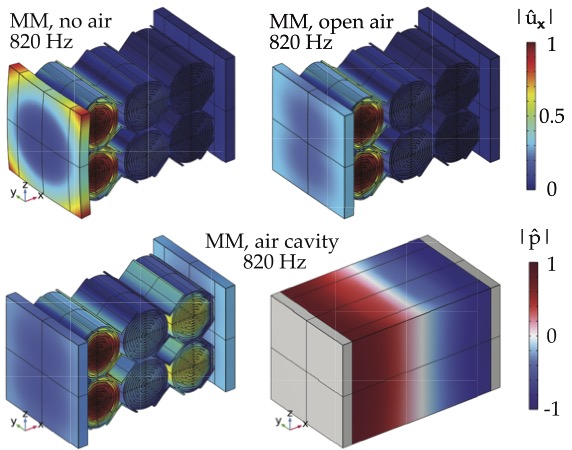}
        \caption{Normalised displacements and pressure}
        \label{subfig:TLsVibroacb}
        \end{subfigure}
     \caption{Transmission loss at normal incidence for the metamaterial specimen (MM) and the benchmark cases (BM) (a) and illustration of the normalised displacement in $x$~direction $|\hat{u}_x|$ and the normalised pressure amplitude $|\hat{p}|$ for three simulated metamaterial cases at \SI{820}{\hertz}.}
     \label{fig:TLsVibroac}
\end{figure}

To further investigate the metamaterial's performances, the transmission loss at normal incidence of the metamaterial specimen and the benchmark cases is presented in Fig.~\ref{subfig:TLsVibroaca}. In contrast to the transfer function, resonances are characterised by a drop of the transmission loss. 
The displacement of the specimen in $x$~direction is illustrated for the purely mechanical case and the case with a closed air cavity in Fig.~\ref{subfig:TLsVibroacb}. By comparing the cases, the movement induced by the air in the second plate is visible. The pressure in the cavity is also illustrated and shows an acoustic resonance mode. 
The transmission loss of the two gypsum plates with an enclosed air volume in the middle (BM1) increases with the frequency but shows a drop at at \SI{106.3}{\hertz} which corresponds to the mass-air-mass resonance of double panels at $f = \frac{1}{2 \pi} \sqrt{2 \rho_0 c^2/d m'}$ where $d$ is the cavity size and $m' = \rho \, t$ is the surface mass density of a plate. BM1 is an effective solution for sound insulation, but also an idealised academic case as discussed in section 3~(c). 
The results for the double panel, in which the gypsum plates are connected by wood studs (BM2), are characterised by a highly fluctuating transmission loss. This behaviour is caused by plate resonances. The metamaterial performs significantly better than BM2 for frequencies above \SI{1100}{\hertz} in all of the three simulated cases. In the lower frequency range, the metamaterial shows a more robust behaviour than the double panel, as there is a high transmission of vibrations for the anti-resonances in the transmission loss of the latter. Due to the robustness at low frequencies and the better performance at higher frequencies, we consider the metamaterial solution a better result than the double plate configuration.
Compared to the concrete wall (BM3), the metamaterial achieves a better performance starting from \SI{520}{\hertz}. In the case of a closed air cavity, besides for BM3 for which the transmission loss of the metamaterial drops below the TL of the latter in two narrow frequency ranges.
The vibro-acoustic simulations show, that a closed air cavity decreases the performance of the metamaterial. The decrease is due to the resonances, especially in the case of a closed air cavity, the movement of the plates which protrude beyond the metamaterial and the interaction between the metamaterial and the air. Additional design parameters have hence to be considered to generate a situation where the response of the specimen is mainly driven by the mechanics.

\subsection{Enhancing the vibro-acoustic performance of the specimen}
\begin{figure}[!h]
    \centering
      \begin{subfigure}[c]{0.32\textwidth}
        \centering
    \includegraphics[trim = 0 0 15 0, clip,width=\textwidth]{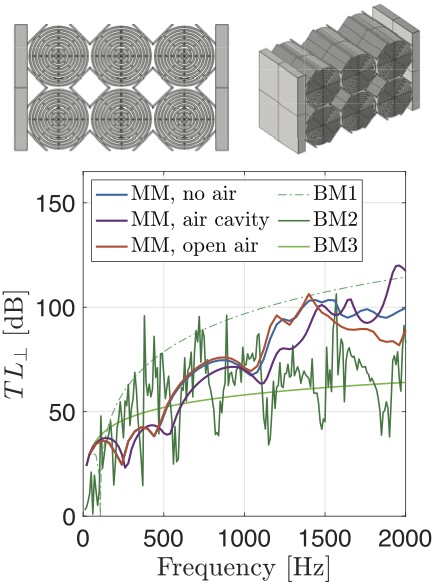}
        \caption{$\alpha$ cut}
        \label{subfig:cutAlpha}
        \end{subfigure}
        \hfill
      \begin{subfigure}[c]{0.32\textwidth}
        \centering
    \includegraphics[trim = 0 0 15 0, clip,width=\textwidth]{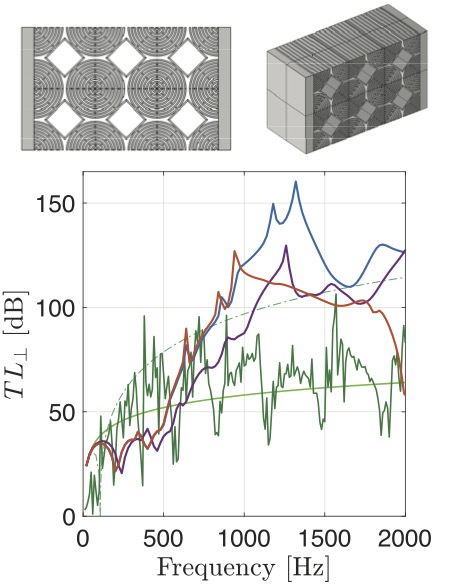}
        \caption{$\beta$ cut}
        \label{subfig:cutBeta}
        \end{subfigure}
        \hfill
      \begin{subfigure}[c]{0.32\textwidth}
        \centering
    \includegraphics[trim = 0 0 15 0, clip,width=\textwidth]{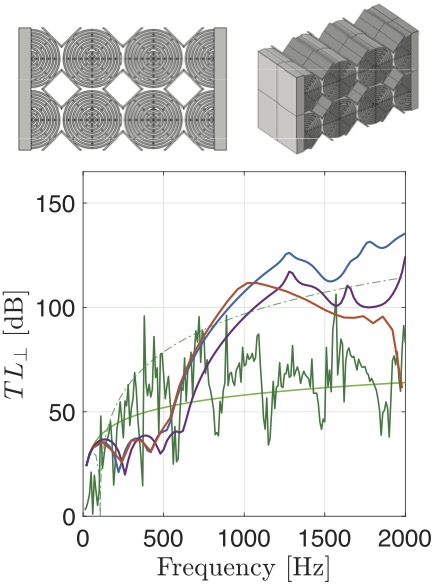}
        \caption{$\gamma$ cut}
        \label{subfig:cutGamma}
        \end{subfigure} 
     \caption{Transmission loss at normal incidence for metamaterial specimens with the same number of unit cells but different interface conditions towards the plates and the air ("cuts").}
     \label{fig:cuts}
\end{figure}

To deploy the metamaterial in an acoustic panel, the vibro-acoustic properties of the specimen have to be improved such that they present an advantage compared to the benchmarks. The first step in reducing the influence of the air is the elimination of direct connections between the plates by the air to avoid cavity resonances, which can be achieved by reducing length and width of the plate (\SI{10.4}{\centi \meter} each) to the dimensions of the metamaterial (\SI{7.3}{\centi \meter}~$\times$~\SI{10}{\centi \meter}). The results and the specimen geometry are shown in Fig.~\ref{subfig:cutAlpha}. 
Compared to the results obtained for the prior design (cf. Fig.~\ref{subfig:TLsVibroaca}) the transmission loss in the purely mechanical case decreases since the corners of the input plate move less with the smaller plates. The apparent behaviour of the structure becomes stiffer, which is observed in a shift of the peaks and drops of the transmission loss to higher frequencies even in the open cavity case. For a closed air cavity the downsizing of the plates improves the transmission loss becoming similar to BM2 at higher frequencies. However, the benchmark cases BM1 and BM2 perform still better than the metamaterial specimen.
In a second step the interface conditions between the metamaterial and plates as well as between metamaterial and the air are modified. In the original design, called "$\alpha$ cut" in the following, an array of three entire unit cells are connected to the plates on four surfaces on each side. In two additional designs, the unit cells are cut in the middle, such that the solid connection area to the plates is maximised (\SI{70}{\percent} instead of \SI{7}{\percent}). Furthermore, the top and bottom surface of the specimen is cut in the same way for the $\beta$ cut (Fig.~\ref{subfig:cutBeta}) while it remained similar to the first case for the $\gamma$ cut (Fig.~\ref{subfig:cutGamma}). 
The results in Fig.~\ref{fig:cuts} show that the performance of the new cuts in the band gap is much improved compared to the performance of the $\alpha$ cut. Besides some peaks in the response of the $\beta$ cut, which are due to local resonances of beams at the boundary and thus enhance the transmission loss, the responses of the new cuts are relatively similar. This result shows, that the interface condition between the metamaterial and the plates plays a crucial role in the performance of the specimens. The higher solid/solid connectivity between metamaterial and plates in new cuts forces the vibration energy to enter into the structure instead of passing mainly into the air.

%%%%%%%%%%%%%%% Discussion %%%%%%%%%%%%%%%
\section{Discussion}

The tested labyrinthine cell has proven to be a good candidate for designing a metamaterial for the mitigation of acoustic and  mechanical vibrations. 
With the enhanced design, the performance of the metamaterial specimen exceeds the practically feasible benchmark cases  starting from \SI{740}{\hertz}. 
To activate the effect of the band gap at lower frequencies, a further vibro-acoustic optimisation of the base unit cell can be performed. Possible steps are a redistribution of the mass inside the unit cell enhancing the local resonances or the addition of heavier materials. 
Another mean to enhance the vibro-acoustic performance of the finite specimens is to increase the number of cells (cf.~Fig.~\ref{subfig:simuDispCurvesTFb}), which leads to a larger frequency range for the band gap. However, a high number of unit cells will be out of the design requirements for interior wall constructions and, furthermore, the lower bound of the band gap found by the Bloch-Floquet analysis cannot be undercut with the same unit. \\
An important point that deserves more attention in future investigations is the importance of the connection between the metamaterial and the continuous plates. It has already been observed, that interface conditions towards a continuous (or Cauchy-like material) change the response of a  metamaterial~\cite{2023_demetriou,2023_perez}, but the significance of the change observed in this study overpassed the expectations. The interface can, thus, be considered as potential optimisation factor for finite metamaterial structures. \\
In the present paper, we validate the performance of the metamaterial at normal incidence against simulations of a double plate with an air cavity, a double panel in which the plates are connected by wood studs and a block of concrete. The next step is the validation in a setting that includes the diffuse-field incidence of a larger-scale panel, which can be obtained experimentally and from numerical simulations. In the latter case either previously designed finite structures or periodically repeated unit cells~\cite{2021_gazzola} can be studied.\\
The finite metamaterial specimens, optimised in the present paper, are now ready to be used as basic building blocks in larger scale structures, such as vibro-acoustic panels. By further upscaling the problem and designing civil engineering elements, computational resources will pose a difficulty. Even though the calculation time could be reduced by \SI{30}{\percent} for the finite-size specimens studied in the present paper, by using a mechanical instead of an acoustic loading, the model still contains a number of degrees of freedom in the order of millions that need to be resolved. 
Furthermore, the design of a larger-scale soundproof panel can include special arrangements of metamaterials that we cannot possibly simulate with the periodicity approach presented in many papers. 
Given the complexity of the involved microstructure and the large dimensions which are targeted, this upscaling will only be possible through the use of an established homogenised model. 
To this end, we can use homogenisation models based on augmented elasticity descriptions. For dynamic cases, models like the relaxed micromorphic model~\cite{2022_voss,2022_demore}, in which the metamaterial is modelled as an equivalent continuum, have proven their validity and can be applied. 
Moreover, the RMM has been recently enhanced to account for surface effects of the type presented in Fig.~\ref{fig:cuts} while remaining in a homogenised framework~\cite{2023b_perez}. 
In this context, it is important that the influence of the air on the performance of the metamaterial remains small to obtain an adequate homogenised model.

%%%%%%%%%%%%%%% Conclusion %%%%%%%%%%%%%%% 
\section{Conclusions}
In this paper, we study a labyrinthine metamaterial with regard to its application for vibro-acoustic control. 
Finite-size specimens are designed by the help of dispersion diagrams including one optimum unit cell design for which the band gap is maximised. 
The specimens are organised as sandwich structures in which the metamaterial is connected to a plate on top and bottom, respectively. 
Three designs with a length of one and three unit cells, respectively, are manufactured and tested experimentally in mechanical vibro-impact tests.
The measurement protocol and the presented post processing methodology allow to obtain an average transfer function of the specimen, similar to a uniform loading, and the sound transmission loss at normal incidence from the experiments. 
The larger samples show a good mechanical performance in attenuating the vibrations for frequencies that lie in the band gap, especially the optimised specimen, and the numerical models could be validated by the experiments. 
Vibro-acoustic numerical simulations include the vibration propagation in the air and reveal a reduction of the sample's performance due to the vibration propagation in the air, which is strongly dependent on the acoustic boundary conditions. 
Numerical studies show, that the vibro-acoustic behaviour of the metamaterial is strongly influenced by the 
solid/solid connection of the metamaterial with a neighbouring homogeneous material. The interface can be considered as an optimisation parameter that enhances the performance of the specimen significantly. 
In perspective, important short term goals are the further optimisation of the specimen's vibro-acoustic properties and investigation on the role of the interface in structures composed from metamaterials and continuous materials. 
A medium term goal is the optimisation of large-scale structures with an optimised distribution of metamaterial units, thanks to the use of the relaxed micromorphic model.
% \vskip6pt

\newpage
%%%%%%%%%%%%%%% Rest %%%%%%%%%%%%%%% 
\dataccess{The frequency response functions, obtained from the experimental point-wise excitation tests, as well as the average accelerations and loads obtained from the numerical simulations are provided.}
\conflict{The authors declare no potential conflicts of interest with respect to the research, authorship, and publication of this article.}
%\funding{This work was supported by the European Commission with the ERC Consolidator Grant META-LEGO [N$^\text{o}$ 101001759].}
\ack{This work was supported by the European Commission with the ERC Consolidator Grant META-LEGO [N$^\text{o}$ 101001759].}

%%%%%%%%%% Insert bibliography here %%%%%%%%%%%%%%
\bibliographystyle{unsrt}
\bibliography{RSTA-2003-0367}

%%%%%%%%%% Insert appendix here %%%%%%%%%%%%%%
\newpage
\appendix

\section{Bloch-Floquet simulations: band gap maximisation study} \label{app:BFsimu}
\begin{figure}[!h]
    \centering
                 \includegraphics[trim = 0 0 0 20, clip,width=0.7\textwidth]{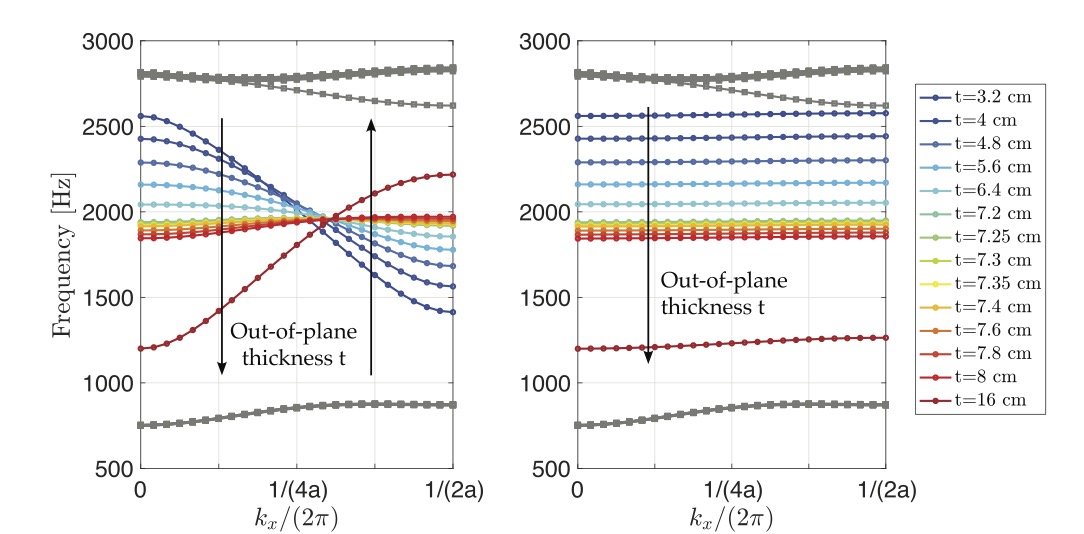}
             \caption{Dispersion curves obtained for the photopolymer for modes 5 (left) and 6 (right), the dispersion curves of modes 4 and 7 obtained for all tested thicknesses are added in grey.}
                \label{fig:modes56optim}
\end{figure}

As depicted in Fig.~\ref{fig:dispCurves}, the 3D Bloch-Floquet simulations reveal that certain modes of the unit cell lead to dispersion curves that invade the band gap. These modes involve out-of-plane movements that are sensitive to the thickness of the unit cell. In a parametric study, the thickness of the photopolymer specimen was varied between \SI{3.2}{\centi \meter} and \SI{16}{\centi \meter} and the evolution of the two modes that invade the band gap are analysed. The results in Fig.~\ref{fig:modes56optim} show the dispersion curves obtained for the two modes as coloured lines.
To highlight the variation of the band gap, the dispersion curves of the prior and the next mode in grey. The modes represented by the grey curves converge to the 2D plane-strain case for all the considered thicknesses. 
The dispersion curve of the fifth mode seems to rotate around a pole point and passes from a negative group velocity to a positive group velocity with increasing thickness. The dispersion curve of the sixth mode has a group velocity of almost zero and the frequency of this mode smoothly decreases with the thickness of the unit cell.

\begin{figure}[!bp]
    \centering
                 \includegraphics[trim = 75 0 30 0, clip,width=0.65\textwidth]{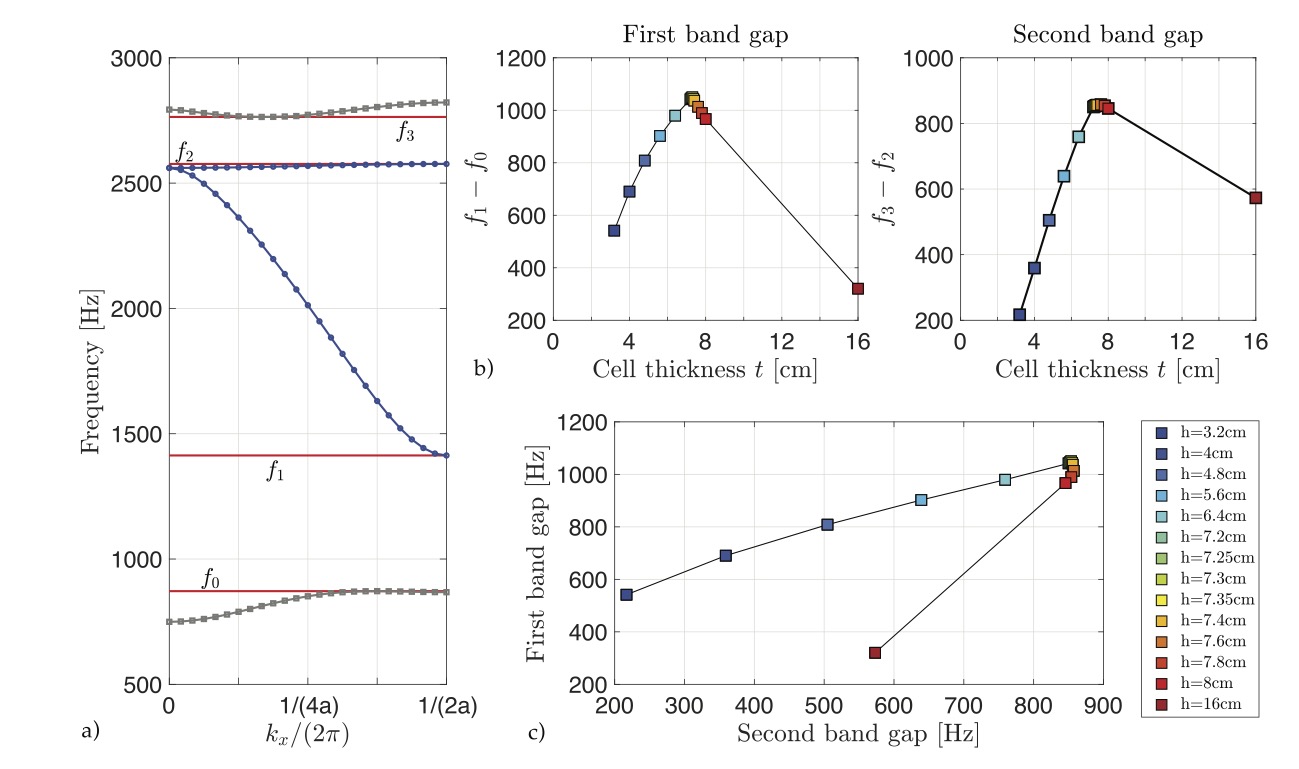}
             \caption{Visualisation of the criteria used to optimise the out-of-plane thickness: Dispersion diagram for one thickness with the characteristic frequencies $f_0$ to $f_3$ (a), band gap size for different thicknesses (b) and confrontation of the two frequency ranges in a single graphic (c).}
                \label{fig:modes56optim2}
\end{figure}
The band gap is maximised for a thickness of \SI{73}{\milli \meter}. The goal of the study was to minimise the frequency range in which the dispersion curves of modes 5 and 6 invade the band gap so as to achieve an almost plane-strain situation also for the 3D specimen. Fig.~\ref{fig:modes56optim2} illustrates the approach that was used to reach this goal. The invading curves divide the band gap in two band gaps, described by the intervals $f_1-f_0$ and $f_3-f_2$ respectively (see Fig.~\ref{fig:modes56optim2}a). The frequency ranges are calculated for each thickness. The optimum thickness maximises both of the criteria, which is illustrated in Fig.~\ref{fig:modes56optim2}b and Fig.~\ref{fig:modes56optim2}c. For the given material parameters and unit cell, the optimum thickness of \SI{7.3}{\centi \meter} was found.

\section{Convergence of the 3D cases to 2D plane-strain cases} \label{app:Converge}
\begin{figure}[!h]
    \centering
    \includegraphics[trim = 0 0 0 0, clip,width=\textwidth]{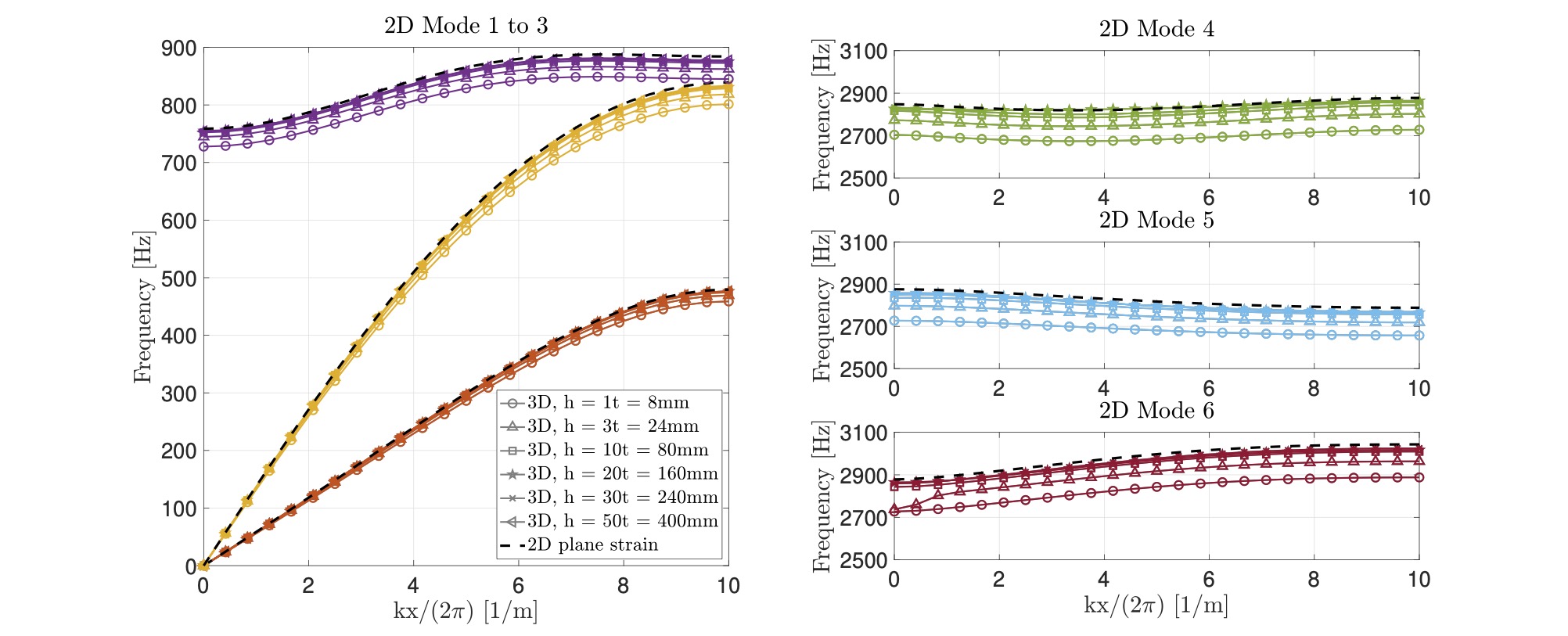}
    \caption{Convergence of the dispersion curves obtained from a 3D simulation case to the 2D plane-strain case for modes 1 to 3 (a) and modes 7 (b), 8 (c) and 9 (d) respectively.}
    \label{fig:freeZ_variationHConv}
\end{figure}

The plane-strain simulation supposes an infinite thickness of the unit cell. In order to test the convergence of the six first in-plane modes in the 3D model, the thickness of the unit cell is gradually increased. Fig.~\ref{fig:freeZ_variationHConv} shows, that the dispersion curves of the 3D model converge to the relations obtained from the plane-strain 2D model when increasing the out-of-plane thickness $t$.

\section{Geometric details of the specimens} \label{app:GeomDetails}
\begin{figure}[!h]
             \centering             
             \begin{subfigure}[c]{0.29\textwidth}
                 \centering
                  \includegraphics[trim = 40 0 80 0, clip,width=\textwidth]{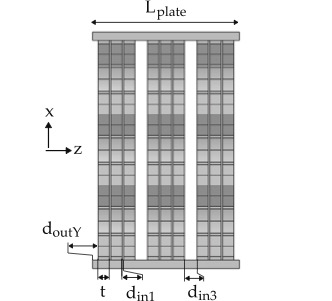}
                 \caption{Side view $x-z$}
                 \label{subfig:specimenSide2annot}
             \end{subfigure}        
             \hfill
             \begin{subfigure}[c]{0.3\textwidth}
                 \centering
                  \includegraphics[trim = 80 0 80 0, clip,width=\textwidth]{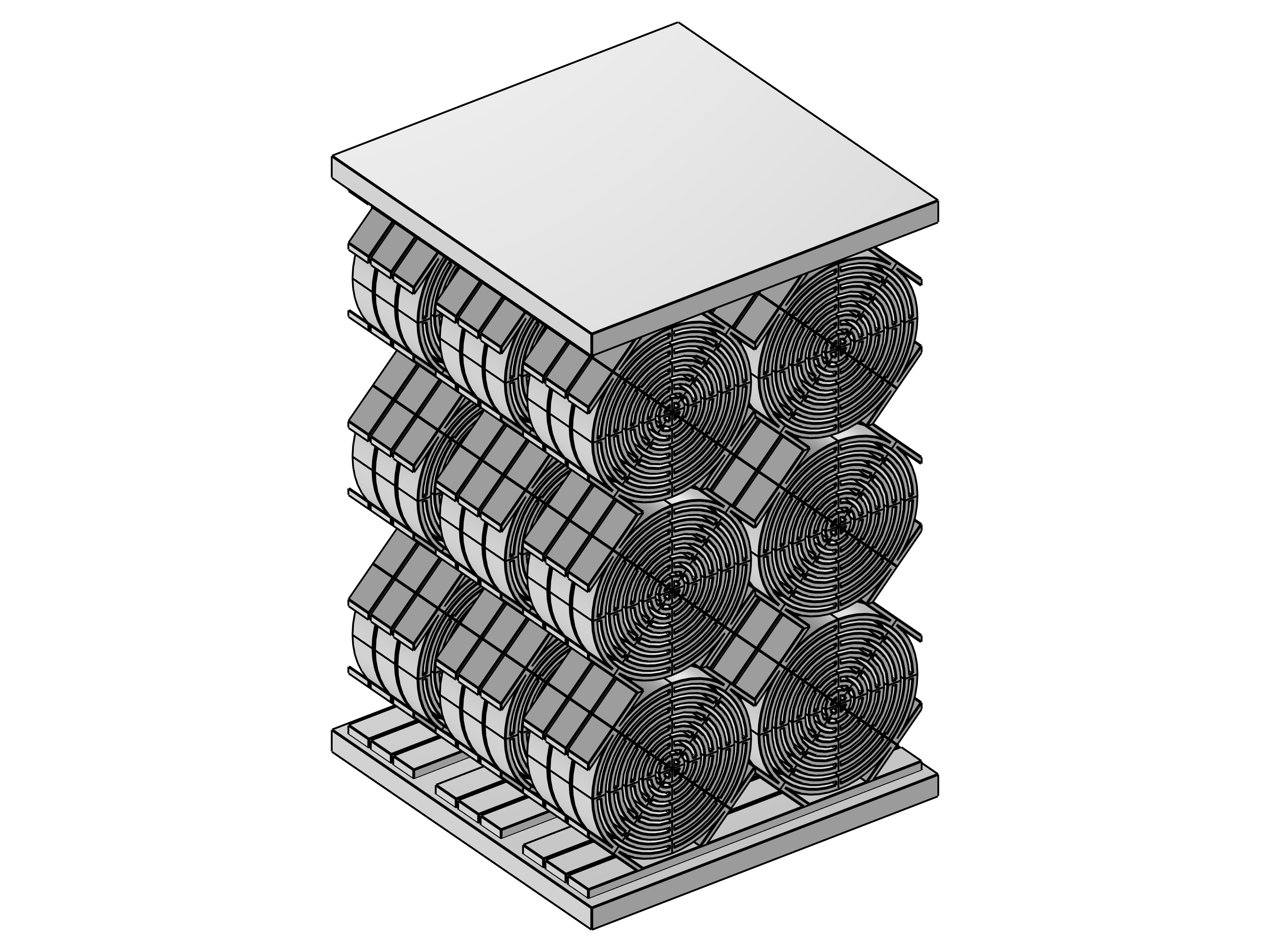}
                 \caption{Perspective view}
                 \label{subfig:specimenPersp}
             \end{subfigure}
             \hfill             
             \begin{subfigure}[c]{0.36\textwidth}
                 \centering
                  \includegraphics[trim = 23 0 65 0, clip,width=\textwidth]{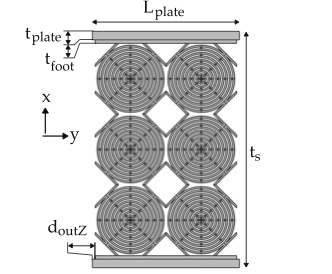}
                 \caption{Side view $x-y$}
                 \label{subfig:specimenSide1annot}
             \end{subfigure}
             \caption{Schematic representation of a specimen illustrating the geometric parameters listed in Tab.~\ref{tab:specDims}.}
             \label{fig:specimenViews}
        \end{figure} 
\begin{table}[!h]
    \centering
    \caption{Summary of the main characteristics of the manufactured specimens.}
    \label{tab:specDims}
    \begin{tabular}{lcccccc}
        \hline
         Specimen name       & sP & bP  & sPG & bPG & s3D & b3D\\
        \hline
        Plate material       & PMMA & PMMA & Gypsum & Gypsum & Photopolymer & Photopolymer\\
        $L_{plate}$    [mm] & 104 & 104  & 104  & 104 & 96 & 104 \\
        $t_{plate}$ [mm] & 6 & 6 & 12.5 & 12.5 & 8 & 8 \\
        Metamaterial        & PMMA & PMMA & PMMA & PMMA & Photopolymer & Photopolymer\\
        $a$ [mm]      & 50 & 50 & 50 & 50 & 50 & 50 \\
        $t$ [mm]  & 8 & 8 & 8 & 8 & 73 & 73 \\        
        Array size        & 2  $\times$  1 & 2  $\times$  3 & 2  $\times$  1 & 2  $\times$  3 & 2  $\times$  1 & 2  $\times$  3 \\
        Array number  & 9 & 9 & 9 & 9 & 1 & 1 \\
        $t_{foot}$ [mm]  & 2.5 & 2.5& 2.5& 2.5 & - & - \\
        $t_s$ [mm] & 67 & 167 & 80 & 180 & 70 & 170 \\
        $d_{outY}$ [mm]  & 4 & 4& 4&4 & 12 & 15.5 \\
        $d_{outZ}$ [mm]  & 2& 2&2 &2 & 12.7 & 16.7\\
        $d_{in3}$ [mm] & 9 & 9& 9& 9& -& -\\
        $d_{in1}$ [mm] & 1 & 1& 1& 1&- &- \\
        \hline
    \end{tabular}
\end{table}%%%End of the table

\section{Derivation of the expressions of the sound pressure} \label{app:deriv}

\subsection{Input plate of the specimen}
Definition of incoming force $F$ as a function of the incident pressure wave $p^{in}$, the reflected pressure wave $p^{ref}$ and the surface $S$ to obtain en expression for $p^{ref}$:
\begin{align}
        &F = (p^{in}+p^{ref}) \, S\\
        &p^{ref} = \frac{F}{S} -p^{in}
\end{align}
Derivation of an expression for the incident pressure of a plane wave:
\begin{equation}
    \begin{aligned}
        \rho_0 \overline{A}^{\,in}(x) &= -\text{grad} \left((p^{in}-p^{ref})\text{e}^{-\mi kx} \right) &&=  -\text{grad}\left( \left( p^{in}- \left(  \frac{F}{S} -p^{in} \right) \right)  \text{e}^{-\mi kx}  \right)\\
        &=  -\text{grad}\left( \left( 2 p^{in}  -\frac{F}{S} \right) \text{e}^{-\mi kx} \right) &&=  -\frac{\partial \left( 2 p^{in}  -\frac{F}{S} \right) \text{e}^{-\mi kx}}{\partial x} \\
        &=  - (- \mi k )\left( \left( 2 p^{in}  -\frac{F}{S} \right) \text{e}^{-\mi kx} \right) &&=   \mi k \left( \left( 2 p^{in}  -\frac{F}{S} \right) \text{e}^{-\mi kx} \right) \\
    \end{aligned}
    \label{eq:developP}
\end{equation}
The incident pressure on the input plate (assuming $x=0$) is obtained from
\begin{align}
    &\rho_0 \overline{A}^{\,in} =   \mi k  \left( 2 p^{in}  -\frac{F}{S} \right) \, ,\\
    &p^{in}  =  \frac{\overline{A}^{\, in}}{2} \, \left( \frac{\rho_0}{\mi k} +\frac{1}{\overline{FRF}^{in}  S}  \right) \, .
    \label{equation:pinc}
\end{align}

\subsection{Output plate of the specimen}

On the output side, only the pressure wave radiating from the specimen is considered. The same development as in  Eq.~\ref{eq:developP} leads to

\begin{align}
    &\rho_0 \overline{A}^{\,out} =    \mi k  \, p^{out} \, , \\
    &p^{out} =  \overline{A}^{\, out} \frac{\rho_0}{\mi k} \,.
    \label{equation:ptra} 
\end{align}

\newpage
\section{Point-wise vs average loading - a numerical study} \label{app:pointwiseVsAverage}

\begin{figure}[!h]
    \centering
      \begin{subfigure}[c]{0.3\textwidth}
         \centering
         \includegraphics[trim = 0 0 0 0, clip,width=\textwidth]{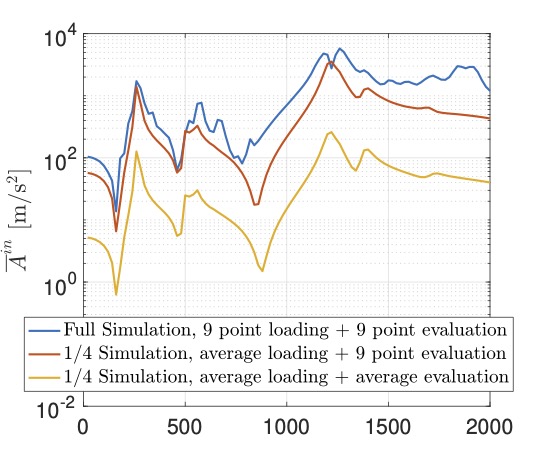}
         \caption{Acceleration input plate}
         \label{subfig:explainDiffsEval1}         
     \end{subfigure}
     \hfill
     \begin{subfigure}[c]{0.3\textwidth}
         \centering
         \includegraphics[trim = 0 0 0 0, clip,width=\textwidth]{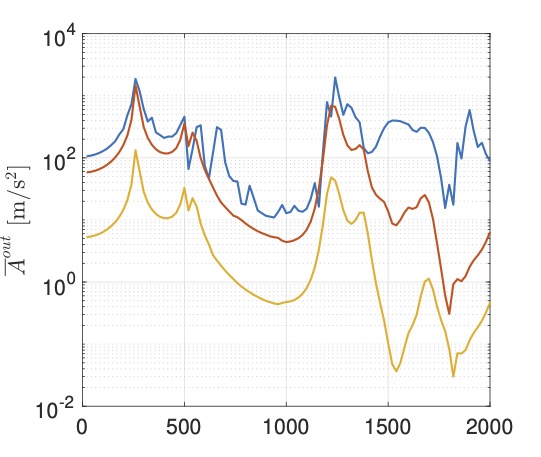}
         \caption{Acceleration output plate}
         \label{subfig:explainDiffsEval2}         
     \end{subfigure}
     \hfill
     \begin{subfigure}[c]{0.3\textwidth}
         \centering
         \includegraphics[trim = 0 0 0 0, clip,width=\textwidth]{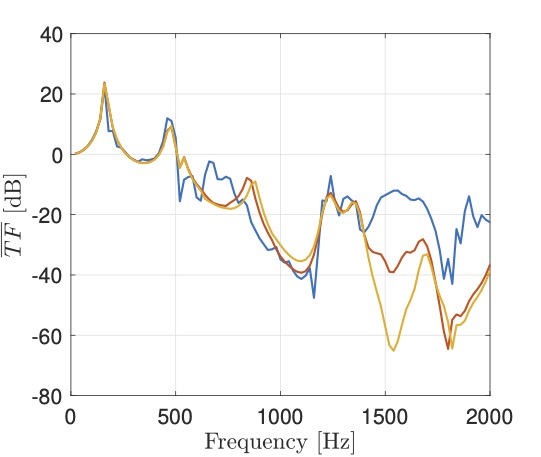}
         \caption{Average transfer function}
         \label{subfig:explainDiffsEval3}         
     \end{subfigure}
     \hfill
    \caption{Example for illustration: influence of loading type and evaluation method on the average acceleration and the transfer function.}
    \label{fig:explainDiffsEval}
\end{figure}

\begin{figure}[!h]
    \centering
     \includegraphics[trim = 98 0 100 0, clip,width=0.8\textwidth]{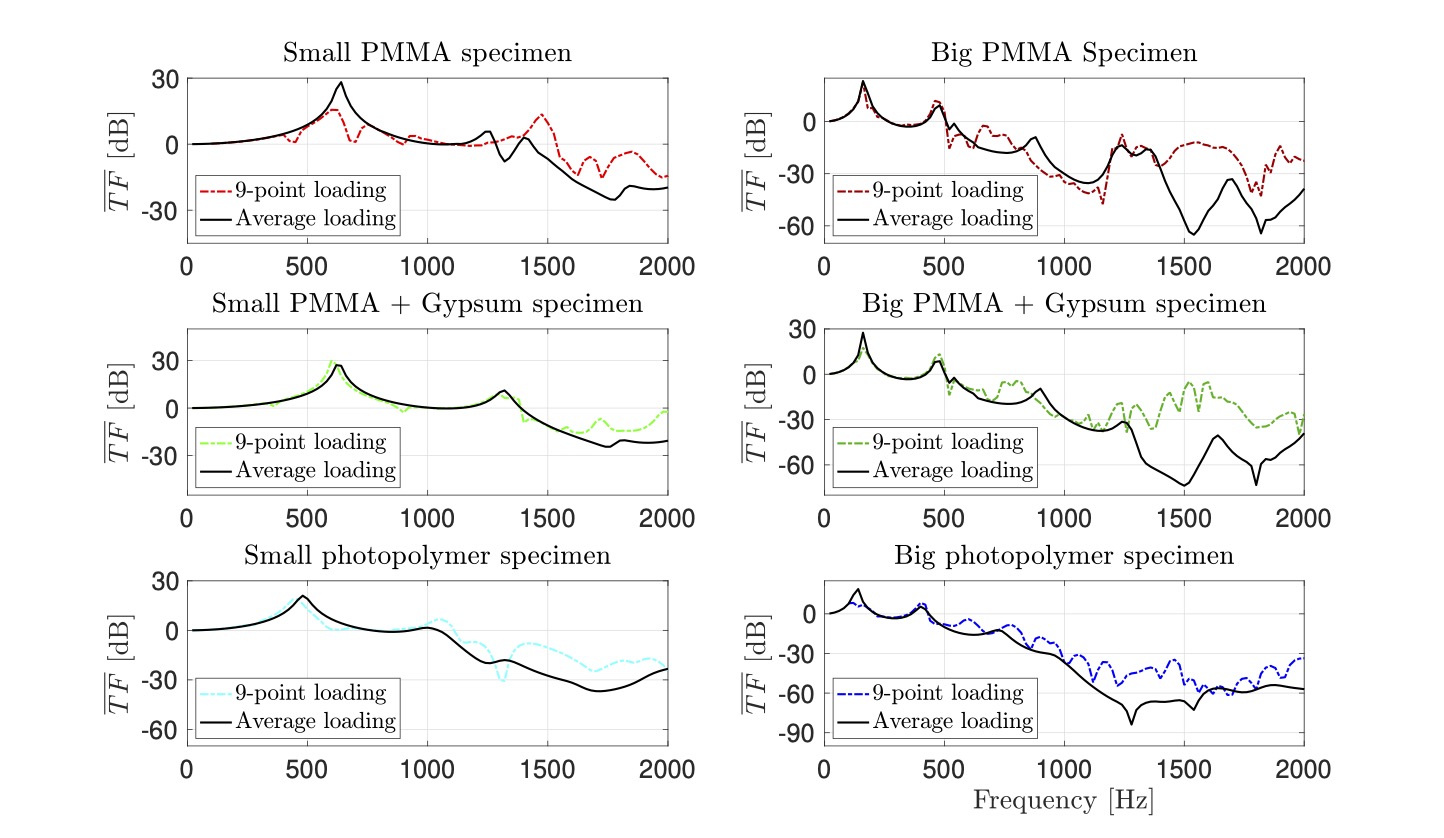}
     \caption{Comparison of 9-point loading and average loading for the 6 tested specimens.}
    \label{fig:simuComparedSimuType}
\end{figure}

In the example depicted in Fig.~\ref{fig:explainDiffsEval}, three different test cases with the big PMMA specimen of this study have been performed in an FE simulation: loading on 9 different points of the specimen input side, evaluation of the acceleration on 9 points on the input side and 9 points on the output side (blue curve), uniform pressure loading of the specimen input side, evaluation of the acceleration on 9 points on the input side and 9 points on the output side (orange curve) and uniform pressure loading of the specimen input side, evaluation of the acceleration on the entire surface on output and input side (yellow curve).
The average accelerations on the top and bottom plate obtained from the three different cases are shown in Fig.~\ref{subfig:explainDiffsEval1} and~\ref{subfig:explainDiffsEval2} respectively. 
The transfer function in Fig.~\ref{subfig:explainDiffsEval3} reveals, that a deviation from the desired result (yellow curve) is generated from the 9-point evaluation (difference yellow and orange curve) as well as from the 9-point loading (difference orange and blue curve), especially at high frequencies. The mismatch at high frequencies indicates that more than 9 points are needed to achieve a better prediction due to more complex movements of the plate. 
However, the transfer functions show a good agreement for most of the frequency range of interest which shows that the method can be applied. A comparison between 9-point loading and average loading obtained from numerical simulations for the experimental specimens is shown in Fig.~\ref{fig:simuComparedSimuType}.

 \end{document}